\documentclass[useAMS,usenatbib,a4paper]{mn2e}
\usepackage{times}
\usepackage{graphicx}
\usepackage{amssymb}
\title[Compact stellar systems in Perseus]{Hubble Space Telescope survey of the Perseus Cluster -IV: Compact stellar systems in the Perseus Cluster core and Ultra Compact Dwarf formation in star forming filaments}
\author[S. J. Penny et al.]{Samantha~J.~Penny$^{1,2,3}$, Duncan~A.~Forbes$^{3}$, Christopher~J.~Conselice$^{4}$\  \footnotemark[0]\\
$^1$School of Physics, Monash University, Clayton, Victoria 3800, Australia\\ 
$^2$Monash Centre for Astrophysics, Monash University, Clayton, Victoria 3800, Australia\\
$^3$Centre for Astrophysics and Supercomputing, Swinburne University of Technology, Hawthorn, Victoria 3122, Australia\\
$^4$School of Physics \& Astronomy, University of Nottingham, Nottingham, NG7 2RD, United Kingdom\\
}
\begin{document}

\maketitle

\begin{abstract}
We present the results of the first search for Ultra Compact Dwarfs (UCDs) in the Perseus Cluster core, including the region of the cluster around the unusual Brightest Cluster Galaxy (BCG) NGC 1275. Utilising \textit{Hubble Space Telescope} Advanced Camera for Surveys imaging, we identify a sample of 84 UCD candidates with half-light radii 10 pc $< r_{e} < 57$ pc out to a distance of 250 kpc from the cluster centre, covering a total survey area of $\sim70$ armin$^{2}$. All UCDs in Perseus lie in the same size-luminosity locus seen for confirmed UCDs in other regions of the local Universe. The majority of UCDs are brighter than $M_{R} = -10.5$, and lie on an extrapolation of the red sequence followed by the Perseus Cluster dwarf elliptical population to fainter magnitudes. However, three UCD candidates in the vicinity of NGC 1275 are very blue, with colours $(B-R)_{0} < 0.6$ implying a cessation of star formation within the past 100 Myr. Furthermore, large blue star clusters embedded in the star forming filaments are highly indicative that both proto-globular clusters (GCs) and proto-UCDs are actively forming at the present day in Perseus.  We therefore suggest star forming filaments as a formation site for some UCDs, with searches necessary in other low redshift analogues of NGC 1275 necessary to test this hypothesis. We also suggest that tidal disruption of dwarf galaxies is another formation channel for UCD formation in the core of Perseus as tidal disruption is ongoing in this region as evidenced by shells around NGC 1275. Finally, UCDs may simply be massive GCs based on strong similarities in the colour trends of the two populations.   
\end{abstract}

\section{Introduction}

Deep, space-based imaging using the \textit{Hubble Space Telescope} has made it possible to study the cores of nearby rich clusters in incredible detail, including their compact stellar systems. The class of compact stellar systems includes globular clusters (GCs) and compact ellipticals (cEs) but had no comparable objects of intermediate luminosity or size, until the discovery of Ultra Compact Dwarfs (UCDs) by \citet{hilker99} and \citet{drinkwater00}. UCDs are typically defined to have sizes 10 $< r_{e} <$ 100 pc and  luminosities of --10.5 $>$ M$_V$ $>$ --14, whereas GCs have lower luminosities and near constant sizes of R$_h$ $\sim$ 3 pc and the proto-type cE M32 has R$_h$ $\sim$ 120 pc and M$_V$ $\sim$ --16.5 \citep{kent87}. 

As sample sizes of compact stellar systems has increased and selection criteria relaxed, it has become clear that the boundaries between between UCDs, GCs and other star clusters has become increasingly blurred \citep{price09,madrid10,forbes11b,brodie11,huxor11}. Thus it is important to fully sample the parameter space of size and luminosity (within a given surface brightness limit) to obtain a true census of the compact stellar systems in any cluster environment.

The origin and nature of UCDs is yet to be determined, with numerous hypotheses put forward to explain the origin of these objects. The simplest explanation for the origin of UCDs is that they are an extension of globular clusters to higher magnitudes and sizes. Evidence for this exists in their colours, with UCDs lying along the red and blue subpopulations of the globular cluster colour-magnitude trend. These objects also follow a similar spatial distribution as the GC system of their host galaxy \citep{mieske08,gregg09}. The larger masses of UCDs might be reached via the mergers of GCs \citep{fellhauer02,kisslerpatig06}.  Therefore uncovering the true nature of UCDs requires not only the study of these bright, massive star clusters, but also the GC system of the environment in which they reside. 

Given UCDs are primarily found in relatively high density regions of the Universe such as clusters, they might be formed during mergers between gas rich galaxies \citep{fellhauer02,bruns11}. When these galaxies interact, colliding gas clouds will be compressed enough to form stars in superclusters with masses $10^{7} - 10^{8}$ M$_{\odot}$ with sizes $\sim100$ pc. Some of these superclusters will undergo mergers, resulting in long-lived, compact objects that survive to the present day as UCDs. 

Alternatively, it has been hypothesised that UCDs could be the nuclear star clusters of dwarf galaxies stripped via tidal interactions \citep{bekki01,drinkwater03}. Due to the dense nature of clusters, encounters between dwarf and massive galaxies would be expected to be commonplace. After several passages around their host cluster or galaxy, the dwarf may be completely stripped of its stellar envelope, with only the dense nucleus remaining. The cores of rich clusters are extremely high density environments, where the strong cluster tidal potential would be hazardous to any object not sufficiently massive to withstand its influence \citep{penny09}.

Uncovering the origin of these objects requires searches in all environments, from isolated galaxies through to the cores of the densest galaxy clusters.  The identification of UCDs is confined to the Local Universe ($D < 100$ Mpc), and must be carried out with high resolution, space-based imaging due to the small physical and, therefore, small angular sizes of these objects. Compact stellar systems have been studied in many nearby clusters of galaxies. These include: Virgo at 16.5 Mpc \citep{hasegan05,evstigneeva07,brodie11},  Fornax at 19 Mpc \citep{hilker99,drinkwater00,gregg09}, Antlia at 35 Mpc \citep{smithcastelli08}, Centaurus at 45 Mpc \citep{misgeld09,mieske09}, Hydra I at 47 Mpc \citep{wehner08,misgeld11b} and Coma at 100 Mpc \citep{price09,madrid10, chiboucas11}.  A notable exception is the Perseus cluster at a distance of 71 Mpc.

This is the first search for UCDs in the Perseus Cluster. At a redshift $v = 5366$ kms$^{-1}$ \citep{Stublerood99}, and distance $D = 71$ Mpc, the Perseus Cluster (Abell 426), is one of the nearest rich galaxy clusters, some 30 Mpc closer that the more frequently studied Coma Cluster. Perseus has a very low spiral galaxy fraction in its core, with the majority of its members being elliptical and S0 galaxies. Given its proximity and richness, it represents an ideal location to identify compact stellar objects.  However, due to its low Galactic latitude ($b$ = --13$^{\circ}$),  it has not been studied in as much detail as other clusters such as Virgo and Coma. 

The Perseus Cluster is furthermore an interesting target for the identification of UCDs due to its highly unusual brightest cluster galaxy (BCG) NGC 1275. NGC 1275 is clearly not a simple system, and exhibits a highly complex structure, including an extended system of filaments. Along the line of sight, there are two systems  \citep{minkowski57}: the high velocity system at $V = 8200$ km s$^{-1}$ falling towards the centre of the cluster, and a low velocity system at $V = 5200$ km s$^{-1}$. A massive, extended system of emission line nebulae extends from the nucleus of the low velocity system.  A number of filaments are seen that  contain young, blue star forming regions, which deserve investigation as sites of ongoing massive star cluster formation at the present time in the Local Universe. 

We carry out a search and analysis of compact stellar systems in the Perseus Cluster of galaxies, utilising high resolution \textit{Hubble Space Telescope (HST)} Advanced Camera for Surveys (ACS) Wide Field  Channel (WFC) imaging in the F555W and F814W bands. In addition to this data, we also use archival ACS data in the F435W and F625W bands to search for UCDs in the vicinity of the cluster centred galaxy NGC 1275. 

In \S~\ref{sec:data} we describe our data set, with the analysis described in  \S~\ref{sec:analysis}. The sizes, colour and properties of the UCDs we identify in our imaging are presented in \S~\ref{sec:ucds}. We examine the GC populations of the BCG NGC 1275 and the giant elliptical NGC 1272 in \S~\ref{sec:gcs}. A discussion of our results including formation scenarios for these objects is presented in \S~\ref{sec:discussion}, with our conclusions in \S~\ref{sec:conclude}.  

Throughout this paper we adopt $D = 71$ Mpc as the distance to the Perseus Cluster, corresponding to an angular scale of $1'' =0.36$~kpc.  All the UCDs and GCs we identify in Perseus are at this time candidates as we do not yet have spectroscopic redshifts to confirm their cluster membership. All sizes presented in the paper for UCDs and GCs are effective (half light) radii.

\section{Data}
\label{sec:data}

\subsection{ACS imaging}

We have obtained \textit{HST} ACS WFC imaging in the F555W and F814W bands for five fields in the core of the Perseus Cluster. The exposure times were one orbit ($\sim2200$~s) per band per field.  The scale of these images is 0.05 arcsec pixel$^{-1}$, and the ACS WFC field of view is $202 \times 202$ arcsec$^{2}$. This results in a survey area of $\sim57$ arcmin$^{2}$ in $V$ and $I$, out to a distance of $\sim250$ kpc from the cluster centre (taken to be NGC 1275). These observations cover the giant elliptical NGC 1272 along with 4 other fields, but exclude the cluster centred galaxy NGC 1275. Six fields were targeted originally, but due to a guide star acquisition failure, observations of one of the fields located at $\alpha =$ 03:19:25.95, $\delta =$ +41:25:33.9 (J2000.0) were not possible.

\begin{figure}
\includegraphics[width=0.47\textwidth]{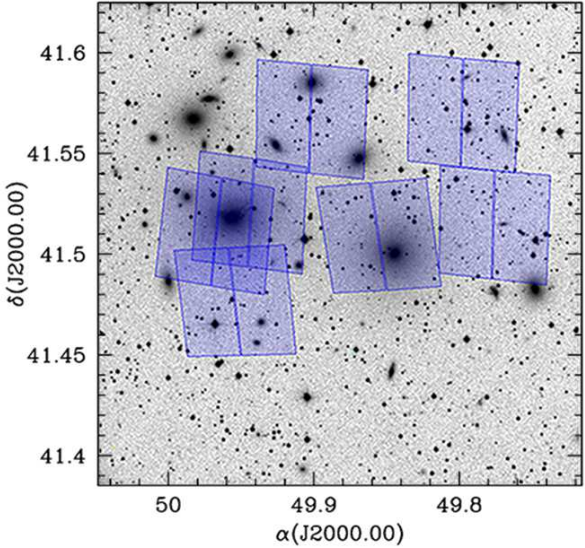}
\caption{The regions of the Perseus Cluster core covered in this study. The highlighted regions represent the 7 ACS WFC fields ($202'' \times 202''$). The two brightest galaxies are visible to the centre left (NGC 1275) and centre right (NGC 1272) of the image. NGC 1275 itself is covered by two overlapping pointings. The survey area for the rest of the cluster is not contiguous, but was rather designed to target dwarf ellipticals visible in ground based imaging whilst avoiding very bright stars.}
\label{surv}
\end{figure}

Data covering the cluster-centred galaxy NGC 1275 (Perseus A) is taken from the $HST$ data archive. The imaging was obtained in the F435W and F625W bands, corresponding to $B$ and $R$ respectively. The total exposure times used in this paper are  4962s in $R$, and 4917s in $B$. These observations are also drizzled to a scale of 0.05 arcsec pixel$^{-1}$. The resulting 7 ACS fields that were successfully targeted are shown in Fig.~\ref{surv}.

\begin{table}
\caption{The 7 $HST$ ACS observations used in this paper. Their positions are shown in Fig.~\ref{surv}. Due to guide star acquisition failure, no data is available for the field Perseus Cluster 2.}
\begin{center}
\begin{tabular}{lccc}
\hline
Field & $\alpha$ & $\delta$ & Filters \\
& (J2000.00) & (J2000.00) &  \\
\hline
Perseus Cluster 1 & 03:19:47.90 & +41:28:00.2 & F555W, F814W  \\
Perseus Cluster 3 & 03:19:23.17 & +41:29:59.0 &  F555W, F814W  \\
Perseus Cluster 4 & 03:19:04.72 & +41:30:10.8 &  F555W, F814W  \\
Perseus Cluster 5 & 03:19:34.57 & +41:33:37.1 & F555W, F814W  \\
Perseus Cluster 6 & 03:19:09.13 & +41:33:49.9 &  F555W, F814W  \\
NGC 1275 SE & 03:19:51.01 & +41:30:41.7 & F435W, F625W \\
NGC 1275 SW & 03:19:45.82 & +41:30:41.5 & F435W, F625W \\
\hline
\end{tabular}
\end{center}
\label{obs}
\end{table}

\section{Data analysis}
\label{sec:analysis}

\subsection{Object identification}

We identify objects in our imaging using \textsc{sextractor} \citep{bertin}. Bright ellipticals were subtracted from our images prior to the object detection. Elliptical isophote fitting of these galaxies were done using the \textsc{iraf} task \textsc{ellipse}. Model galaxies were then created using the \textsc{iraf} routine \textsc{bmodel}, with the task \textsc{imarith} used to subtract these models from the original imaging. This subtraction reveals a large compact star cluster population which would otherwise have been missed during object identification via \textsc{sextractor}. This population may include potential UCD candidates.

The complex nature of NGC 1275 makes the subtraction of this galaxy via ellipse fitting very difficult.  The presence of a saturated nucleus, dust lanes and filaments  resulted in a poor subtraction of the underlying galaxy, with the dust lanes obscuring many globular clusters in these central regions. Likewise, removing the galaxy's light via the subtraction of a smoothed version of the image (unsharp masking) would leave the extensive dust lanes and other substructure behind, obscuring a large central region of the galaxy. The presence of dust would furthermore lead to large uncertainties in the colours of central star clusters. Therefore we decided not to remove this galaxy from the original image. As a result, the study of this particular region of the cluster excludes the inner 6~kpc region of NGC 1275. 

Objects were identified in the galaxy subtracted images using \textsc{sextractor}, with photometry performed on the original, non galaxy subtracted images. We identified all objects at least 5 contiguous pixels in size at 3$\sigma$ above the noise level independently in both bands, and then matched the catalogues. This catalogue matching ensures that objects detected in only one band (e.g. cosmic rays that were not properly cleaned) do not make it into our final catalogue, as would be the case if detection was carried out in one band only.

\subsection{Photometry}

The total magnitudes are measured using \textsc{sextractor isocor}. The flux of the object above a threshold level is measured, and then a correction is made for the flux lost in the wings of the object, assuming the profile of the object is Gaussian. The background for each object is estimated locally as it varies across the field of view  due to the presence of giant ellipticals and intra-cluster light. Colours are measured using fixed 3 pixel apertures centred on each object to ensure the same extent of the objects in both bands are being sampled in the colour calculation.   The \textit{F555W} and \textit{F814W} magnitudes are then converted to \textit{V} and \textit{I} respectively using the transformations of \citet{sirianni}. Transformations are likewise applied to the  \textit{F435W} and  \textit{F625W} magnitudes to convert them to $B$ and $R$. 

Reddening corrections are then applied to our photometry. The Perseus Cluster is at a low Galactic latitude, and lies in a region of variable extinction \citep{burstein84}. To check for any effects that this might have on the measured colours of our objects, we find the reddening corrections at the centres of each of our \textit{HST} pointings using the extinction maps of \citet{schlegel}. The values of $E(B-V)$ vary between 0.160 and 0.164 across our survey area, therefore we choose not to vary our reddening corrections with cluster position. Our reddening corrections do not take into account internal extinction by NGC 1275, which contains patchy dust. The likely effect of reddening caused by the dust of  NGC 1275 is discussed in \S~\ref{sec:ucd1275}. A colour magnitude diagram constructed from the initial \textsc{sextractor} catalogues reveals a large amount of foreground and background contamination, with a sequence of objects belonging to Perseus also visible. 
  
\subsection{Size measurements}

Sizes for all objects were determined using the fitting code \textsc{ishape} \citep{larsen09}, which is commonly used to derive the shape parameters of objects that are only slightly resolved such as extragalactic globular clusters and UCDs (e.g. \citealt{harris09,madrid10}). \textsc{ishape} requires a model of the PSF, and we describe the fitting of this PSF below. 

\subsubsection{Point spread function fitting}

In order to determine the sizes of our objects usign \textsc{ishape}, we need to build the ACS PSF. Given the low Galactic latitude of Perseus ($b$ = -13$^{\circ}$), numerous suitable stars are available in each field for which to fit the PSF, therefore we build a separate PSF for each frame. The ACS PSF is built using the \textsc{iraf} package \textsc{daophot}. The \textsc{daophot} task \textsc{pstselect} was used to identify stars from which to construct the \textit{ACS} PSF.  Bright but non saturated stars that are well isolated were then selected manually from the list provided by \textsc{pstselect} with which to build the PSF. The PSF was computed with the task \textsc{psf}, and then oversampled by a factor of 10 as required by \textsc{ishape}. 

\subsubsection{Size determination}

The PSF is then used in the determination of the sizes of our objects using \textsc{ishape}. The effective radius, $r_{e}$ is determined for each object by fitting a PSF convolved profile to each object in the image. By varying the full width at half maximum (FWHM) of the model profile, \textsc{ishape} finds a best fit to the object using the given analytical model. We fit all objects using a King profile with a concentration parameter $c = 30$, as previous authors have found this to be the typical concentration parameter for globular clusters and UCDs across a range of locations. The fitted FWHM is then converted to an effective radius (containing half the object's light) using the conversion $r_{e} = 1.48\times$~FWHM provided in the \textsc{ishape} documentation for a King profile with a concentration parameter $c = 30$. 

The fitting is performed on the original, non galaxy subtracted imaging, and is effective down to $\sim10\%$ of the FWHM of the PSF \citep{harris09b}. This corresponds to 0.2 pixels for ACS/WFC, which is 3.5 pc at the distance of Perseus. Objects must therefore have effective radii larger than 3.5 pc to be included in our final sample of star clusters with reliable size measurements. However, all objects with  $r_{e}>1$ pc (i.e. non-zero radii, or not stars) and S/N$>10$ in both bands are included in plots showing both UCDs and GCs. All sizes presented in the paper for UCDs and GCs are effective radii. 

There is a large foreground contamination of stars which need to be removed from our sample that at fainter magnitudes are visually indistinguishable from globular clusters and UCDs. \textsc{ishape} typically fits stars with effective radii of zero, allowing for their easy removal from our sample. However, saturated stars will be fit with non-zero sizes by \textsc{ishape}, with flat-topped light distributions that do not follow the PSF. It is therefore essential that these objects are cleaned from our catalogue. Due to the fact that a point source (i.e. star) may saturate in one band but not the other, it would therefore be fit by \textsc{ishape} with a non-zero FWHM in the saturated band. It follows that the radius of each object is checked in both the F814W and F555W bands to ensure our catalogues are properly cleaned of point sources. We therefore remove all objects that are saturated in at least one band. This check is likewise performed for the F435W and F625W imaging of the NGC 1275 region.

We make no selection cuts based on colour, size or ellipticity prior to running \textsc{ishape} on the sample. Instead, our catalogues were cleaned via visual inspection of all objects after the \textsc{ishape} fitting routine was run to remove spurious detections from our sample. Obvious bright background galaxies, edge objects, dwarf ellipticals, stars, and other artefacts (such as diffraction spikes) are then removed  by eye from the catalogue of objects successfully fitted by \textsc{ishape}.   

\begin{figure*}
\includegraphics[width=0.97\textwidth]{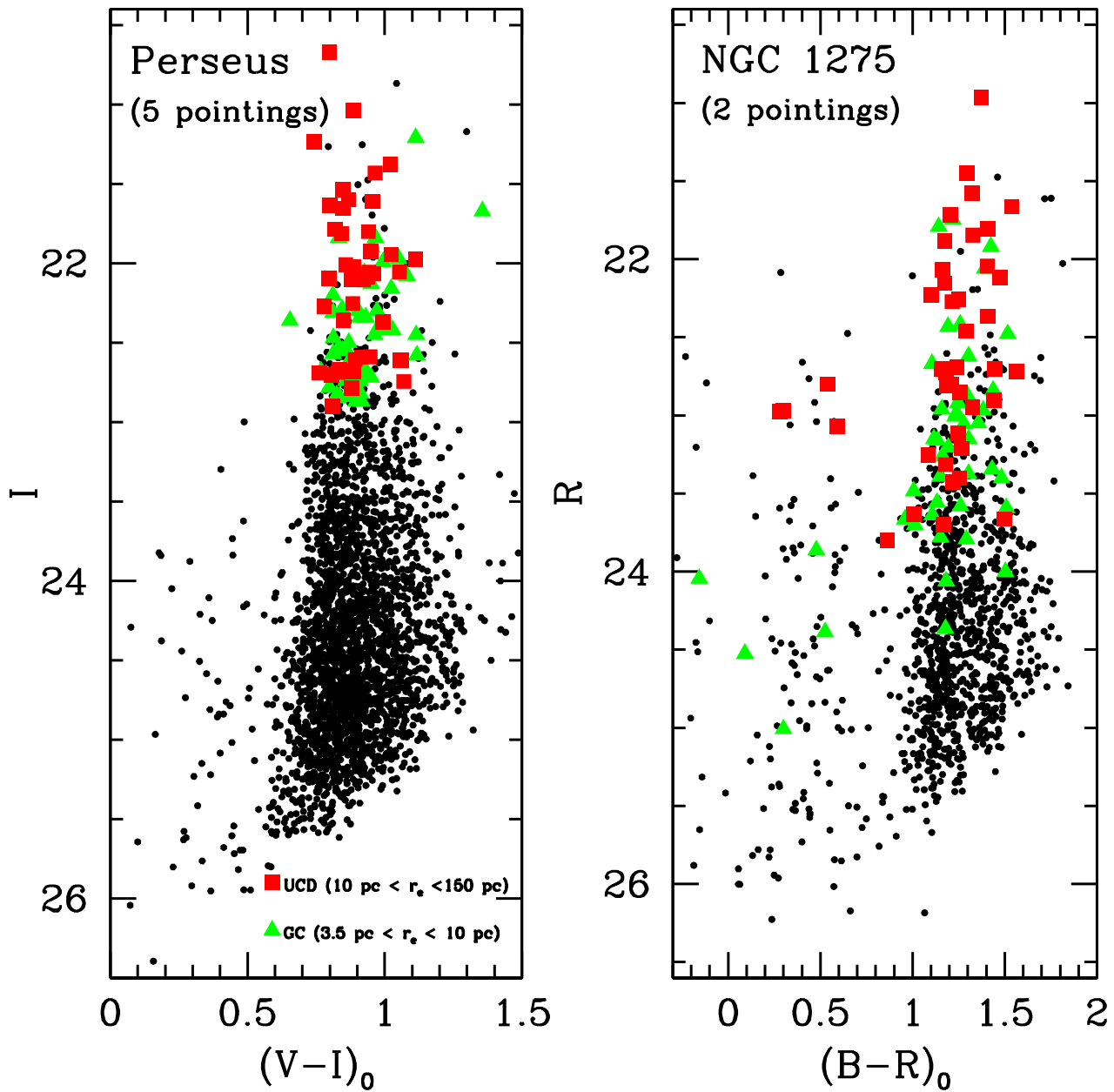}
\caption{Colour magnitude diagrams for the compact stellar objects around NGC 1275 (right), and the five pointings covering the rest of the Perseus Cluster (left). We have removed stars, background objects, and other contaminants from our sample, but otherwise we include all objects with effective radii 1-150~pc. Sequences defined by the compact stellar systems are visible, with a small additional population of blue $(B-R)_{0} \sim 0.4$ star clusters seen for NGC 1275. Objects with reliable size measurements (S/N $>40$ and $r_{e} < 3.5$~pc) are highlighted, with UCDs denoted by red squares, and GCs by green triangles. All other objects are denoted by black dots, and have insufficient S/N to have accurate sizes determined, else are smaller than 3.5~pc.}
\label{both_cmds}
\end{figure*}

Compact star clusters are then identified in this catalogue. We define compact objects to have sizes 1~pc~$<r_{e} <150$~pc with objects $r_{e}<10$~pc classed as globular clusters, and compact objects larger than this taken to be UCDs. Thus we follow \citet{brodie11} in not imposing a luminosity criteria on UCDs. The upper size limit of 150~pc ensures background ellipticals do not make our final sample, with the largest UCD we identify having size $r_{e} = 57$~pc. The cleaned colour magnitude diagrams for these two regions are shown in Fig.~\ref{both_cmds}. Only compact objects are shown in this plot. This combination of manual cleaning and the removal of stars via PSF fitting using \textsc{ishape} is highly successful, with clear colour sequences defined by the compact stellar systems. 

At the distance of Perseus, a typical globular cluster of $r_{e} = 2.5$~pc will be too small to have its size accurately determined by \textsc{ishape}, as their effective radii will be less than 10\% of the ACS PSF. Therefore any compact objects we identify in this imaging with reliable sizes will be larger than 3.5~pc, and will be extended globular clusters or larger objects. 

\section{Ultra compact dwarfs}
\label{sec:ucds}

Identifying UCDs requires the measurements of their sizes, which necessitates their signal-to-noise to be sufficiently high for \textsc{ishape} to determine accurate sizes. Therefore, to identify the minimum S/N at which sizes for compact objects can be accurately determined, we compare the sizes of objects $>3.5$~pc in both bands as determined by \textsc{ishape} at different S/N thresholds. At low S/N ($<30$), objects are typically not well matched in both bands, with a large scatter between the effective radii measured in the F814W and F555W bands. However, at $S/N > 40$ the sizes of the objects are in general well matched in both bands. We therefore select only those objects with a minimum signal-to-noise of 40 in the F555W band for our UCD identification. which is the shallower of the two bands.  The same S/N criteria are applied to the F435W and F625W band imaging for NGC 1275. 

\begin{figure}
\includegraphics[width=0.47\textwidth]{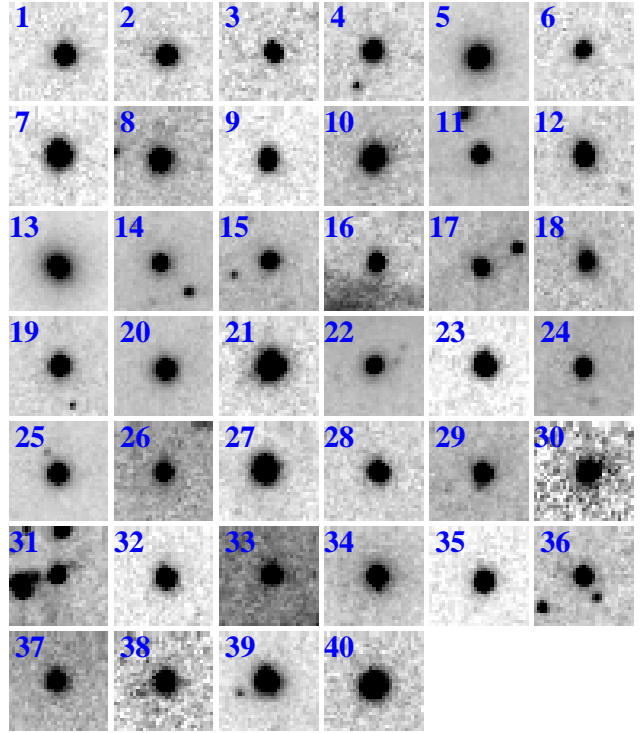}
\caption{F625W band images of 40 candidate UCDs in the vicinity of NGC 1275. The cutouts measure $2'' \times 2''$ (700pc $\times$ 700pc at the distance of Perseus). All objects were selected to have sizes  10~pc~$<r_{e} <150$~pc, though the largest object in our sample has $r_{e} = 57$~pc. The ID numbers in the top left hand corner of each cutout are those given in Table~\ref{tab1275}, which also lists their positions, magnitudes and colours.}
\label{ucds1275}
\end{figure}
 
UCDs typically have effective radii 10~pc~$<r_{e} <150$~pc. To create our sample of candidate UCDs, objects must have sizes that fall in this range in both the \textit{F814W} and \textit{F555W} bands, and a minimum S/N of 40 in both bands. We follow \citet{brodie11} making no colour or luminosity selection criteria. All objects matching these criteria are then carefully examined in both the F555W and F814W imaging, and any objects missed in our initial cleaning with visually clumpy or galaxy-like substructure in either band are removed.  Identical selection criteria are used to identify UCD candidates in the NGC 1275 region.

\begin{figure}
\includegraphics[width=0.47\textwidth]{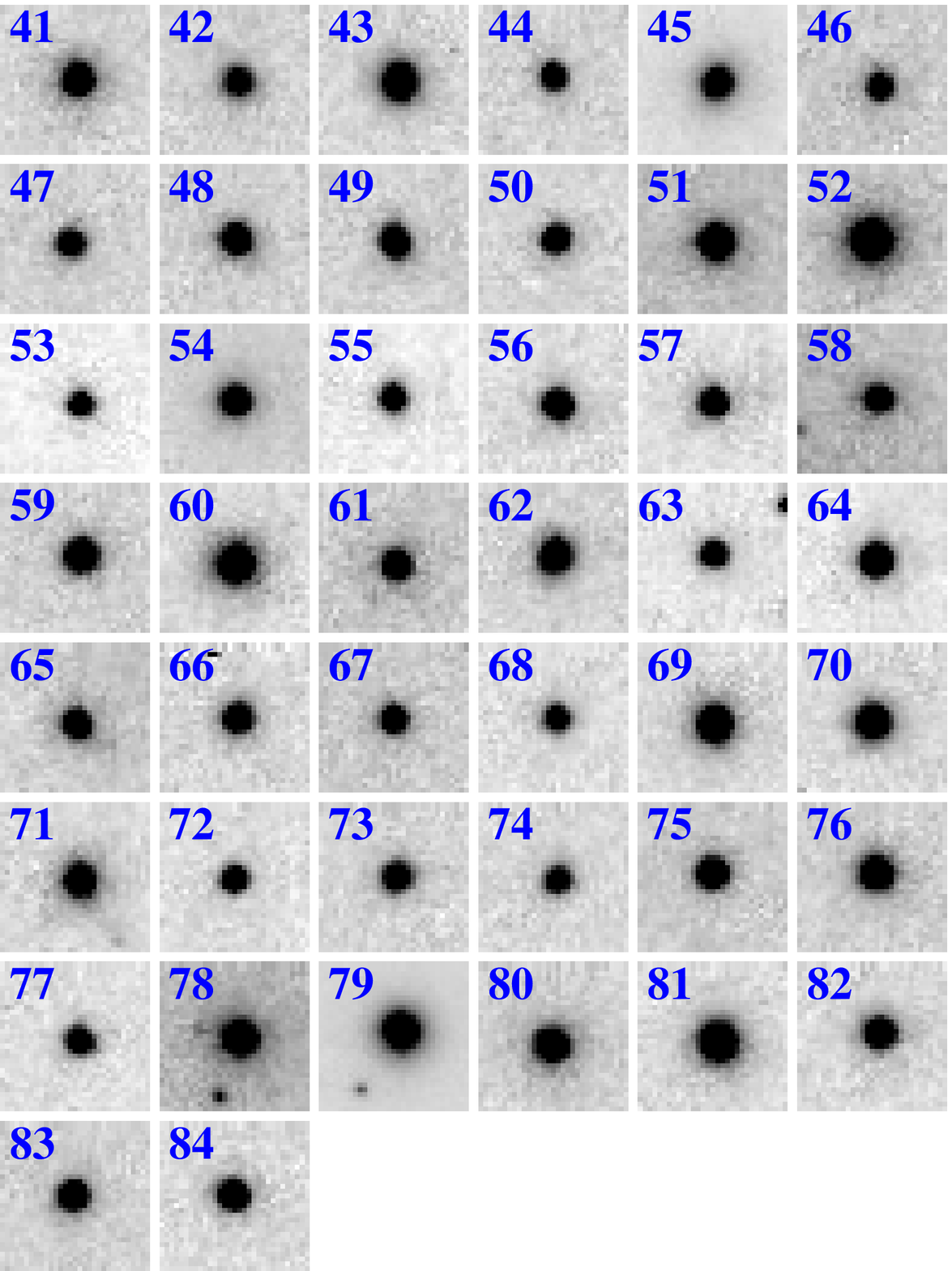}
\caption{F814W images of the remaining 44 UCD candidates in the Perseus Cluster core identified from F555W and F814W imaging. The cutouts measure $2'' \times 2''$ ($700 \times 700$ pc at the distance of Perseus.)  The number in the top left hand corner is the ID number assigned to each UCD in Table~\ref{tabper}. }
\label{vicutouts}
\end{figure}

Cutouts of all candidate UCDs in our imaging are shown in Figs.~\ref{ucds1275} and \ref{vicutouts}, and their properties  listed in Tables~\ref{tab1275} and \ref{tabper}. This imaging shows that all UCDS we identify are round systems, despite no ellipticity selection. We note that some H$\alpha$ filaments are present in the imaging of UCDs around NGC 1275.

\begin{table*}
\caption{UCDs in the vicinity of NGC 1275 identified from F435W and F625W imaging.}\label{tab1275}
\begin{center}
\begin{tabular}{lcccccc}
\hline
UCD & $\alpha$ & $\delta$ & $R$ & $(B-R)_{0}$ & $r_{e,F625W}$ & $b/a$\\
& (J2000.0) & (J2000.0) & (mag) & (mag) & (pc) & \\
\hline
  UCD1  & 03:19:37.64 & +41:29:43.8 & 22.81 $\pm$ 0.02 & 1.18 $\pm$ 0.05 & 11.7 & 0.97\\
  UCD2  & 03:19:39.79 & +41:30:17.2 & 22.90 $\pm$ 0.03 & 1.44 $\pm$ 0.06 & 26.9 & 0.73\\
  UCD3  & 03:19:40.00 & +41:28:58.7 & 23.66 $\pm$ 0.04 & 1.50 $\pm$ 0.08 & 11.2 & 0.62\\
  UCD4  & 03:19:40.16 & +41:30:33.2 & 23.12 $\pm$ 0.03 & 1.25 $\pm$ 0.06 & 18.9 & 0.63\\
  UCD5  & 03:19:40.72 & +41:30:53.4 & 21.45 $\pm$ 0.01 & 1.29 $\pm$ 0.03 & 24.8 & 0.91\\
  UCD6  & 03:19:40.89 & +41:29:59.3 & 23.63 $\pm$ 0.03 & 1.01 $\pm$ 0.07 & 10.4 & 0.61\\
  UCD7  & 03:19:41.03 & +41:32:11.1 & 22.04 $\pm$ 0.02 & 1.41 $\pm$ 0.04 & 12.5 & 0.82\\
  UCD8  & 03:19:41.17 & +41:30:49.2 & 22.70 $\pm$ 0.02 & 1.44 $\pm$ 0.05 & 13.9 & 0.97\\
  UCD9  & 03:19:41.60 & +41:32:07.9 & 22.85 $\pm$ 0.02 & 1.26 $\pm$ 0.05 & 14.4 & 0.60\\
  UCD10 & 03:19:41.66 & +41:30:22.3 & 22.46 $\pm$ 0.02 & 1.29 $\pm$ 0.04 & 17.6 & 0.78\\
  UCD11 & 03:19:44.36 & +41:31:04.3 & 22.27 $\pm$ 0.02 & 1.21 $\pm$ 0.04 & 13.1 & 0.93\\
  UCD12 & 03:19:44.64 & +41:28:59.6 & 22.70 $\pm$ 0.02 & 1.16 $\pm$ 0.05 & 20.8 & 0.75\\
  UCD13 & 03:19:45.13 & +41:32:06.0 & 20.97 $\pm$ 0.01 & 1.37 $\pm$ 0.03 & 57.0 & 0.96\\
  UCD14 & 03:19:45.56 & +41:30:43.8 & 22.07 $\pm$ 0.02 & 1.16 $\pm$ 0.04 & 13.1 & 0.99\\
  UCD15 & 03:19:45.95 & +41:30:33.3 & 22.69 $\pm$ 0.02 & 1.24 $\pm$ 0.05 & 13.6 & 0.88\\
  UCD16 & 03:19:48.06 & +41:31:52.0 & 22.26 $\pm$ 0.02 & 1.25 $\pm$ 0.06 & 15.2 & 0.75\\
  UCD17 & 03:19:48.56 & +41:31:17.8 & 22.97 $\pm$ 0.03 & 0.28 $\pm$ 0.04 & 18.4 & 0.84\\
  UCD18 & 03:19:49.11 & +41:29:42.9 & 23.80 $\pm$ 0.04 & 0.86 $\pm$ 0.07 & 33.0 & 0.80\\
  UCD19 & 03:19:49.96 & +41:31:21.6 & 21.88 $\pm$ 0.02 & 1.17 $\pm$ 0.03 & 13.1 & 0.85\\
  UCD20 & 03:19:49.96 & +41:30:10.3 & 21.72 $\pm$ 0.01 & 1.21 $\pm$ 0.03 & 15.7 & 0.92\\
  UCD21 & 03:19:50.05 & +41:32:08.9 & 22.23 $\pm$ 0.02 & 1.10 $\pm$ 0.04 & 31.2 & 0.94\\
  UCD22 & 03:19:50.14 & +41:30:39.9 & 21.81 $\pm$ 0.02 & 1.41 $\pm$ 0.04 & 13.1 & 0.80\\
  UCD23 & 03:19:50.15 & +41:29:14.3 & 22.36 $\pm$ 0.02 & 1.41 $\pm$ 0.05 & 11.5 & 0.75\\
  UCD24 & 03:19:51.32 & +41:30:44.7 & 22.73 $\pm$ 0.02 & 1.18 $\pm$ 0.05 & 19.2 & 0.85\\
  UCD25 & 03:19:51.43 & +41:30:17.3 & 21.66 $\pm$ 0.01 & 1.54 $\pm$ 0.03 & 11.5 & 0.83\\
  UCD26 & 03:19:51.80 & +41:31:33.9 & 23.32 $\pm$ 0.03 & 1.18 $\pm$ 0.06 & 13.9 & 0.76\\
  UCD27 & 03:19:51.80 & +41:29:44.4 & 21.58 $\pm$ 0.01 & 1.32 $\pm$ 0.03 & 16.3 & 0.84\\
  UCD28 & 03:19:52.27 & +41:29:42.3 & 22.95 $\pm$ 0.03 & 1.32 $\pm$ 0.05 & 11.7 & 0.74\\
  UCD29 & 03:19:52.79 & +41:30:16.0 & 22.80 $\pm$ 0.02 & 0.54 $\pm$ 0.04 & 22.4 & 0.78\\
  UCD30 & 03:19:52.83 & +41:29:16.3 & 23.70 $\pm$ 0.04 & 1.17 $\pm$ 0.08 & 14.9 & 0.86\\
  UCD31 & 03:19:52.96 & +41:30:04.6 & 23.04 $\pm$ 0.03 & 0.55 $\pm$ 0.05 & 10.7 & 0.73\\
  UCD32 & 03:19:53.01 & +41:32:25.8 & 23.25 $\pm$ 0.03 & 1.08 $\pm$ 0.06 & 29.8 & 0.64\\
  UCD33 & 03:19:53.10 & +41:30:15.9 & 23.21 $\pm$ 0.03 & 1.27 $\pm$ 0.06 & 10.4 & 0.83\\
  UCD34 & 03:19:53.53 & +41:30:20.4 & 22.15 $\pm$ 0.02 & 1.17 $\pm$ 0.04 & 16.8 & 1.00\\
  UCD35 & 03:19:54.08 & +41:29:26.1 & 22.80 $\pm$ 0.02 & 1.21 $\pm$ 0.05 & 18.6 & 0.67\\
  UCD36 & 03:19:54.98 & +41:30:22.7 & 22.72 $\pm$ 0.02 & 1.56 $\pm$ 0.06 & 39.7 & 0.74\\
  UCD37 & 03:19:57.19 & +41:31:11.1 & 23.43 $\pm$ 0.03 & 1.22 $\pm$ 0.07 & 25.3 & 0.97\\
  UCD38 & 03:19:57.23 & +41:29:19.2 & 23.40 $\pm$ 0.03 & 1.26 $\pm$ 0.07 & 11.5 & 0.80\\
  UCD39 & 03:19:58.08 & +41:32:39.6 & 21.85 $\pm$ 0.02 & 1.33 $\pm$ 0.04 & 22.1 & 0.89\\
  UCD40 & 03:19:58.74 & +41:29:55.1 & 22.12 $\pm$ 0.02 & 1.47 $\pm$ 0.04 & 13.9 & 0.78\\
\hline
\end{tabular}
\end{center}
\end{table*}

\begin{table*}
\caption{UCDs identified from our F555W and F814W band imaging covering five fields of the Perseus core.}\label{tabper}
\begin{center}
\begin{tabular}{lcccccc}
\hline
UCD & $\alpha$ & $\delta$ & $I$ & $(V-I)_{0}$ & $r_{e,F814W}$ & $b/a$\\
& (J2000.0) & (J2000.0) & (mag) & (mag) & (pc) & \\
\hline
  UCD41 & 03:19:00.38 & +41:31:29.6 & 21.80 $\pm$ 0.02 & 0.94 $\pm$ 0.04 & 19.2 & 0.98\\
  UCD42 & 03:19:01.72 & +41:33:25.6 & 22.61 $\pm$ 0.03 & 1.06 $\pm$ 0.06 & 22.9 & 0.90\\
  UCD43 & 03:19:03.03 & +41:30:54.9 & 21.81 $\pm$ 0.02 & 0.84 $\pm$ 0.04 & 33.3 & 0.87\\
  UCD44 & 03:19:04.69 & +41:32:03.8 & 22.69 $\pm$ 0.03 & 0.83 $\pm$ 0.06 & 10.4 & 0.61\\
  UCD45 & 03:19:08.07 & +41:29:16.7 & 20.67 $\pm$ 0.01 & 0.80 $\pm$ 0.03 & 33.8 & 0.85\\
  UCD46 & 03:19:09.73 & +41:28:56.9 & 22.70 $\pm$ 0.03 & 0.81 $\pm$ 0.06 & 11.7 & 0.81\\
  UCD47 & 03:19:11.26 & +41:30:59.4 & 22.68 $\pm$ 0.03 & 0.88 $\pm$ 0.06 & 14.7 & 0.94\\
  UCD48 & 03:19:13.29 & +41:32:47.3 & 22.10 $\pm$ 0.03 & 0.88 $\pm$ 0.05 & 14.1 & 0.82\\
  UCD49 & 03:19:13.33 & +41:31:55.1 & 22.09 $\pm$ 0.03 & 0.79 $\pm$ 0.05 & 16.0 & 0.52\\
  UCD50 & 03:19:13.70 & +41:34:22.8 & 22.36 $\pm$ 0.03 & 0.99 $\pm$ 0.06 & 11.2 & 0.81\\
  UCD51 & 03:19:17.07 & +41:29:24.0 & 21.37 $\pm$ 0.02 & 1.02 $\pm$ 0.04 & 14.4 & 0.91\\
  UCD52 & 03:19:17.24 & +41:32:32.4 & 21.04 $\pm$ 0.02 & 0.88 $\pm$ 0.03 & 22.6 & 0.97\\
  UCD53 & 03:19:20.02 & +41:31:38.6 & 22.79 $\pm$ 0.03 & 0.88 $\pm$ 0.06 & 15.7 & 0.80\\
  UCD54 & 03:19:20.04 & +41:28:51.9 & 21.23 $\pm$ 0.02 & 0.74 $\pm$ 0.03 & 18.1 & 0.96\\
  UCD55 & 03:19:25.21 & +41:31:37.4 & 22.61 $\pm$ 0.03 & 1.06 $\pm$ 0.06 & 18.4 & 0.98\\
  UCD56 & 03:19:25.56 & +41:30:49.6 & 22.02 $\pm$ 0.02 & 0.88 $\pm$ 0.05 & 11.2 & 0.89\\
  UCD57 & 03:19:26.15 & +41:34:29.0 & 22.10 $\pm$ 0.03 & 0.89 $\pm$ 0.05 & 13.3 & 0.85\\
  UCD58 & 03:19:27.57 & +41:32:43.9 & 22.25 $\pm$ 0.03 & 0.88 $\pm$ 0.05 & 13.3 & 0.53\\
  UCD59 & 03:19:28.14 & +41:28:55.9 & 21.94 $\pm$ 0.02 & 1.02 $\pm$ 0.05 & 18.6 & 0.90\\
  UCD60 & 03:19:28.30 & +41:34:30.0 & 21.43 $\pm$ 0.02 & 0.96 $\pm$ 0.04 & 49.0 & 0.95\\
  UCD61 & 03:19:28.37 & +41:29:11.3 & 22.36 $\pm$ 0.03 & 0.85 $\pm$ 0.05 & 24.8 & 0.99\\
  UCD62 & 03:19:29.36 & +41:32:54.5 & 21.97 $\pm$ 0.02 & 1.11 $\pm$ 0.05 & 20.0 & 0.84\\
  UCD63 & 03:19:31.41 & +41:30:21.9 & 22.59 $\pm$ 0.03 & 0.91 $\pm$ 0.06 & 10.9 & 0.82\\
  UCD64 & 03:19:32.83 & +41:34:33.9 & 21.60 $\pm$ 0.02 & 0.87 $\pm$ 0.04 & 13.3 & 0.97\\
  UCD65 & 03:19:33.31 & +41:34:34.3 & 22.05 $\pm$ 0.02 & 1.05 $\pm$ 0.05 & 13.6 & 0.64\\
  UCD66 & 03:19:34.88 & +41:32:28.7 & 22.27 $\pm$ 0.03 & 0.78 $\pm$ 0.05 & 15.7 & 0.88\\
  UCD67 & 03:19:37.95 & +41:32:04.8 & 22.67 $\pm$ 0.03 & 0.83 $\pm$ 0.06 & 14.7 & 1.00\\
  UCD68 & 03:19:38.34 & +41:26:43.2 & 22.74 $\pm$ 0.03 & 1.07 $\pm$ 0.07 & 17.3 & 0.67\\
  UCD69 & 03:19:38.86 & +41:27:42.0 & 21.54 $\pm$ 0.02 & 0.85 $\pm$ 0.04 & 22.1 & 0.89\\
  UCD70 & 03:19:40.47 & +41:28:09.3 & 21.78 $\pm$ 0.02 & 0.82 $\pm$ 0.04 & 20.0 & 0.98\\
  UCD71 & 03:19:40.88 & +41:28:03.7 & 21.93 $\pm$ 0.02 & 0.95 $\pm$ 0.05 & 26.1 & 0.80\\
  UCD72 & 03:19:42.19 & +41:33:59.5 & 22.69 $\pm$ 0.03 & 0.76 $\pm$ 0.06 & 10.1 & 0.99\\
  UCD73 & 03:19:42.26 & +41:28:42.5 & 22.60 $\pm$ 0.03 & 0.89 $\pm$ 0.06 & 21.8 & 0.92\\
  UCD74 & 03:19:42.46 & +41:27:35.5 & 22.90 $\pm$ 0.04 & 0.81 $\pm$ 0.07 & 21.0 & 0.71\\
  UCD75 & 03:19:43.11 & +41:29:05.6 & 22.10 $\pm$ 0.03 & 0.92 $\pm$ 0.05 & 12.3 & 0.89\\
  UCD76 & 03:19:45.82 & +41:28:45.8 & 22.01 $\pm$ 0.02 & 0.86 $\pm$ 0.04 & 24.5 & 0.99\\
  UCD77 & 03:19:46.10 & +41:26:58.5 & 22.59 $\pm$ 0.03 & 0.94 $\pm$ 0.06 & 11.7 & 0.98\\
  UCD78 & 03:19:47.59 & +41:29:25.0 & 21.63 $\pm$ 0.02 & 0.80 $\pm$ 0.04 & 23.4 & 0.94\\
  UCD79 & 03:19:49.63 & +41:28:35.4 & 19.91 $\pm$ 0.01 & 1.10 $\pm$ 0.02 & 27.4 & 0.94\\
  UCD80 & 03:19:50.61 & +41:27:41.8 & 21.61 $\pm$ 0.02 & 0.96 $\pm$ 0.04 & 16.0 & 0.95\\
  UCD81 & 03:19:53.52 & +41:28:04.0 & 21.65 $\pm$ 0.02 & 0.85 $\pm$ 0.04 & 46.1 & 0.88\\
  UCD82 & 03:19:53.54 & +41:28:12.7 & 22.06 $\pm$ 0.02 & 0.96 $\pm$ 0.05 & 17.0 & 0.92\\
  UCD83 & 03:19:53.80 & +41:28:57.6 & 22.08 $\pm$ 0.03 & 0.94 $\pm$ 0.05 & 11.2 & 0.74\\
  UCD84 & 03:19:55.47 & +41:27:54.0 & 22.05 $\pm$ 0.02 & 0.94 $\pm$ 0.05 & 10.9 & 0.79\\
\hline
\end{tabular}
\end{center}
\end{table*}

\subsection{Background contamination}

A possible source for contamination in our sample of Perseus Cluster UCDs is background galaxies. Such objects may appear as round, compact objects with colours and angular sizes similar to UCDs at the distance of Perseus. Therefore to calculate the likely contamination of such objects in our sample, we need to utilise a deep survey that is free of low redshift objects which may have their own UCD/GC systems and with similar filters to the data we have used for our  UCD search in Perseus. 

To estimate the background contamination in our UCD sample, we utilise two fields from the GOODS ACS survey, with one field taken from GOODS-N and another field taken from GOODS-S.  We use Section 33 of the N mosaic, and Section 22 of the S mosaic.  Version 1 of the imaging is used in our analysis, which is of sufficient depth to allow comparison to our imaging.  The survey area covered by each mosaic section is $404'' \times 404''$, approximately 4 times the survey area of a single pointing of ACS.  We therefore utilise two mosaic cutouts to approximate our survey area. The imaging was taken in the $F606W$ and $F775W$ bands, which can be easily converted to $V$ and $I$ using \citet{sirianni}. 

\begin{figure}
\includegraphics[width=0.47\textwidth]{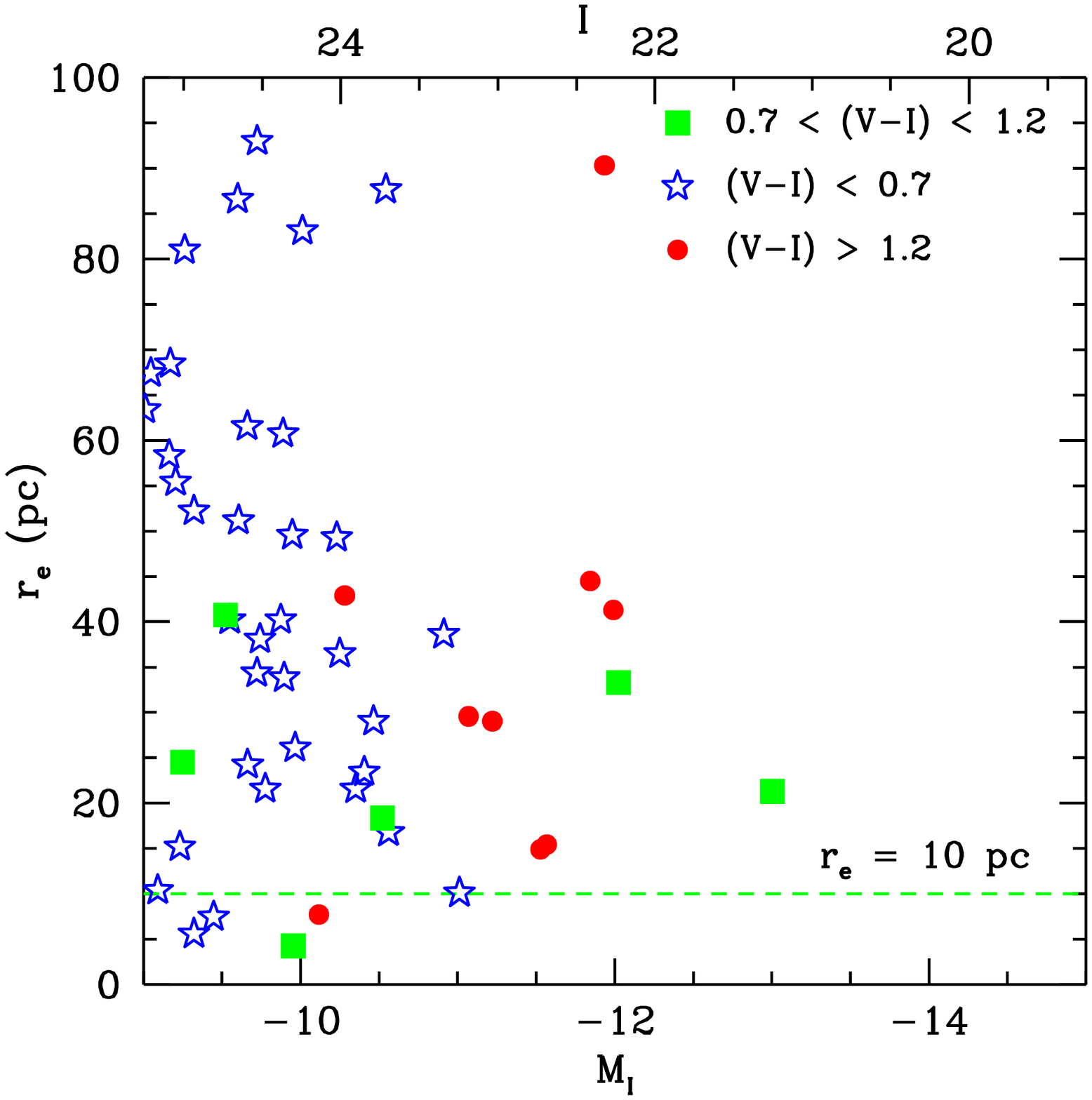}
\caption{The size-magnitude relation for compact objects identified from GOODS ACS imaging, used to estimate the likely contamination of our sample from background galaxies. The squares represent high redshift objects that would be identified as UCDs under our selection criteria and fall in the colour range occupied by ``normal'' Perseus UCD candidates ($0.7< V-I <1.2$). The stars are objects with colours bluer than Perseus UCDs, and red dots are those redder than typical UCD colours.}
\label{bgcheck}
\end{figure}

Objects are identified using the same \textsc{sextractor} code as in our initial object identification. The PSF was then constructed using the \textsc{iraf daophot} task for each mosaic section. Fitting of all objects with \textsc{ishape} was then performed, and the final catalogues cleaned of stars and obvious background galaxies. The sizes, colours  and magnitudes of the remaining objects were then examined in Fig.~\ref{bgcheck}.

In general, the colours of UCD-like objects in the GOODS fields are not consistent with those found for UCD candidates in Perseus ($V-I  \sim 1$). UCD-sized objects in the comparison fields are typically very blue in colour, with $V-I < 0.5$, and are therefore likely strongly star forming, high redshift galaxies. Furthermore, these objects are in general faint, with $I < 23.5$, and the majority of these objects would not be of sufficient signal-to-noise to make the UCD candidate list due to the shallower depth of our imaging. 

All UCD candidates we identify in our F555W and F814W band imaging are brighter than $I = 24$, with colours $0.7 < (V-I) < 1.2$.  For the GOODS-N imaging, only one object has a colour, size and magnitude that falls within the size-colour-magnitude space occupied by ``normal'' Perseus UCDs.  For the GOODS-S field, two such objects are seen. As a result of this exercise, we consider the likely contamination from background sources in our candidate list to be small (2-3 objects). As we have no redshifts at this stage, all UCDs presented in this paper are candidates, with follow-up spectroscopy essential to establish cluster membership. 

For the young, blue UCDs around NGC 1275, the contamination from background objects becomes more of a concern.  Another way to establish the likelihood of cluster membership for compact objects is to examine the radial surface density distribution around their host galaxy. We expect the number of compact systems to fall off with increasing galactocentric distance, before the distribution flattens at large radii to a constant background radii. At large radii the probability of an object being a background contaminant therefore increases. 

\begin{figure}
\includegraphics[width=0.45\textwidth]{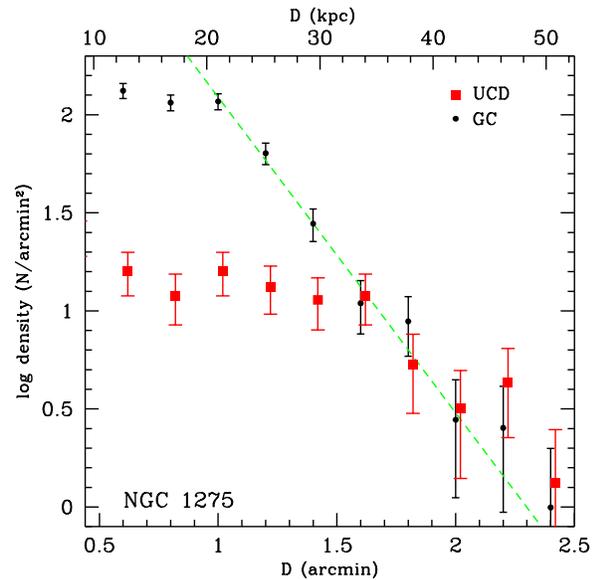}
\caption{Radial surface density of compact objects around NGC 1275. The red squares are the UCDs, with the dots representing GCs. The UCD radial surface densities are arbitrarily shifted to allow comparison between the radial surface density distributions of the UCDs and GCs. A small offset of $+0.02'$ in distance is applied to UCD data set to allow easier comparison of the two populations. The error bars are Poissonian errors on the number counts for each radial bin. The green dashed line is a power law fit to the data of slope $-1.6$. The density for both UCDs and GCs do not flatten out at larger radii, showing that all objects we identify here are statistically associated with NGC 1275, and we have not reached the background level. }
\label{rdist}
\end{figure}

We investigate the surface density distribution of all star clusters in our sample around NGC 1275 in Fig.~\ref{rdist}. The survey region around NGC 1275 is contiguous, with only compact objects in the very inner regions of the galaxy excluded from our search. Therefore we can examine the radial distribution of compact objects in this region of the cluster. The surface density plots for GCs ($r_{e} < 10$~pc) and UCDs (10~pc~$<r_{e} <150$~pc) are both plotted. For GCs, the distribution flattens in the inner regions ($D < 1'$), where a combination of the fact that we were not able to remove the central galaxy and tidal effects start to come into account. A power law of slope $-1.6$ is fit to the GC radial surface density distribution for $D > 1'$. 

The radial density distribution also flattens for UCDs, but at distances $D < 1.5'$ (30~kpc).  This could imply that UCDs are not commonly formed at these galactocentric distances, else large star clusters are more likely to be tidally truncated at small distances from their host galaxy. However, at larger radii both populations follow the same radially decreasing distribution within the error bars. This suggests UCDs and GCs have a similar parent population. This distribution does not flatten off to our largest galactocentric distance, showing the majority of objects in this study are associated with the host galaxy.  

The fact that the radial surface distributions of both GCs and UCDs around NGC 1275 do not flatten off at large radii in this plot, shows that objects in this study are statistically associated with NGC 1275. Therefore given the proximity of the 3 young, blue UCDs to NGC 1275 ($D = 12.8, 21.0, 23.5$~kpc), and the fact that they are located near filaments with ongoing star formation, we consider these likely members of the Perseus Cluster associated with NGC 1275. 

\subsection{UCDs around NGC 1275}
\label{sec:ucd1275}

\begin{figure}
\includegraphics[width=0.47\textwidth]{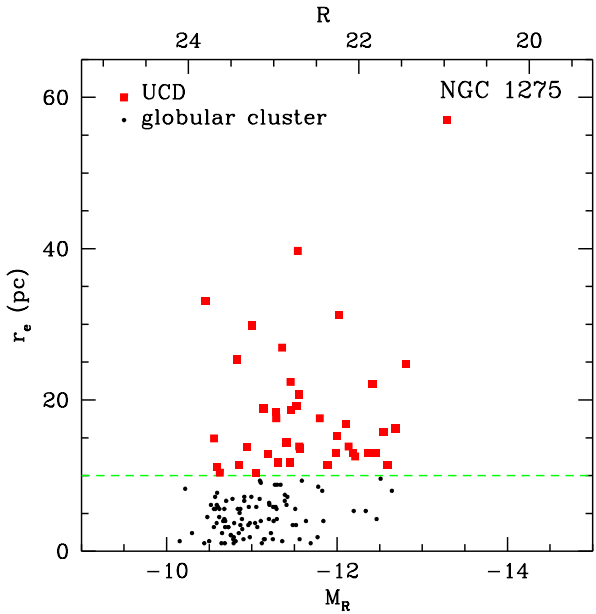}
\caption{Size-magnitude diagram for objects around NGC 1275. The dashed line corresponds to an effective radius of 10 pc, which is  taken to be the size separating the globular cluster and UCD populations. All objects have S/N $> 40$. The red points are UCDs, and the black points are globular clusters.  }
\label{sizes1275}
\end{figure}

\begin{figure}
\begin{center}
\includegraphics[width=0.45\textwidth]{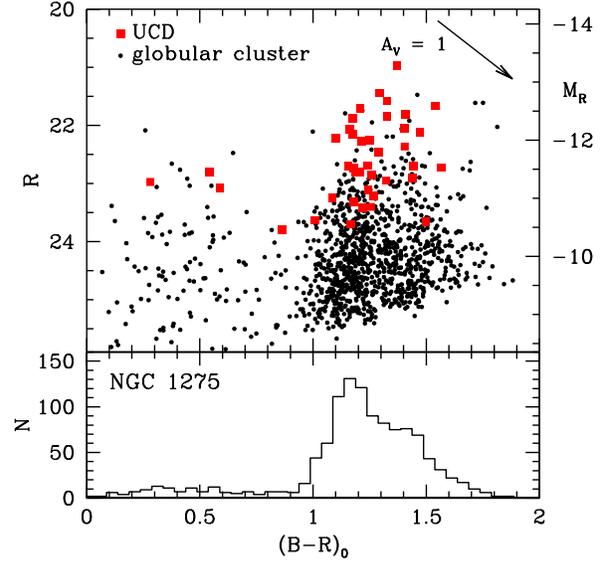}
\caption{The colour distribution of compact stellar systems around NGC 1275. The colours of the globular cluster system appears bimodal, with blue and red subpopulations. The arrow shows the direction and magnitude of the reddening vector. UCDs around NGC 1275 are marked by red boxes. Most NGC 1275 UCDs lie on the bright end of the blue GC subpopulation, with fewer red subpopulation UCDs. A small number of very blue UCDs are seen.}
\label{cols1275}
\end{center}
\end{figure}

To separate UCDs and globular clusters, we plot the effective radius versus absolute magnitude for all objects with S/N$>40$ in F555W in Fig.~\ref{sizes1275}. All objects with sizes $r_{e}> 10$ pc are taken to be UCDs.  Based on their sizes (Fig.~\ref{sizes1275}), we identify a total of 40  UCD sized candidates around the cluster centred galaxy NGC 1275 with sizes 10 pc $ < r_{e} < 57$ pc and magnitudes  between $R = 20.97$ and $R = 23.80$ ($R\approx-14$ to $R\approx-10.5$ respectively). Following the methodology of \citet{brodie11}, we have not restricted the luminosity range for UCDs. We have also made no selection by ellipticity or colour, as not to limit our sample to old, passive systems. The properties of these objects are listed in Table.~\ref{tab1275}, with their colours and magnitudes shown in Fig.~\ref{cols1275}. The UCDs have ellipticity values of ${b/a}$ between 0.6 and 1, consistent with ellipticities found in previous studies \citep{madrid10}, with a mean ratio $b/a = 0.82\pm0.11$.

The majority of these UCDs lie on the same colour magnitude trend followed by the globular clusters in this region. Furthermore, it can be seen that the some UCDs typically follow the ``blue tilt'' seen for the GC blue subpopulation of NGC 1275, with a smaller number of UCDs lying on the red GC subpopulation. 

Unusually, not all of the UCD candidates we identify have colours typically found for UCDs and GCs to date i.e. they do not all lie on a bright extension of the GC colour magnitude trend. Three of these objects (UCD17, UCD29 and UCD31) are bluer than typical UCDs, with $(B-R)_{0}$ colours of 0.28, 0.54 and 0.59 respectively. We refer to these objects as blue UCDs from here onwards.  Another object, UCD18, has a colour of $(B-R)_{0}=0.86$, intermediate between a blue star forming object and a normal UCD. In addition to these UCDs, a number of round objects  are embedded in the star forming filaments that are likely young GCs and UCDs, and can be seen in Fig.~\ref{both_cmds}. However, these regions are too crowded for accurate sizes to be determined for these objects using \textsc{ishape}. They lie in the magnitude range $-8.26 < M_{R} < -12.25$, and their colours are in the range $-0.2 < (B-R)_0 <  0.71$, again consistent with a population that has recently ceased star formation.

\begin{figure}
\includegraphics[width=0.47\textwidth]{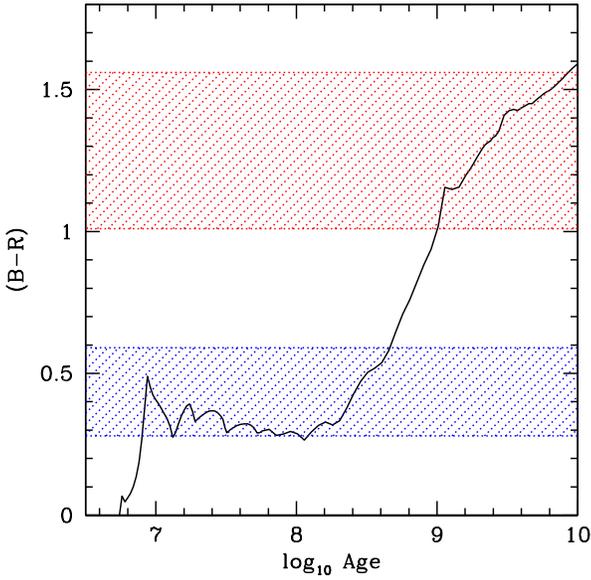}
\caption{Colour-age predictions for the UCDs around NGC 1275. The colours of the NGC 1275 UCDs are compared  to the \citet{bruzualcharlot03} models for how colour evolves with age, using a Saltpeter IMF. The model is shown as a solid line, corresponding to a metallicity of $0.02 (Z_{\odot})$.  It can be seen that the very blue UCDs have ages consistent with a recent cessation of star formation, having young ages of $10$-$100$ Myr. The UCDs with ``typical'' colours are consistent with no recent star formation, and ages $>1$~Gyr.}
\label{colev}
\end{figure}

Similar to \citet{canning10}, we examine the ages of the NGC 1275 UCDs by comparing their colours to \citet{bruzualcharlot03} models for the colour evolution of a simple stellar population since the last burst of star formation (Fig.~\ref{colev}). The colours of the normal UCDs i.e. those with $(B-R)_{0} > 1.0$ are consistent with objects that have not undergone star formation in the past  Gyr. However, the UCD candidates with colours bluer than this are likely much younger objects, with the blue object at $(B-R)_{0} = 0.28$ having a stellar age $\sim100$ Myr. This is consistent with the results of \citet{canning10} for blue star clusters in this region.

\begin{figure}
\includegraphics[width=0.47\textwidth]{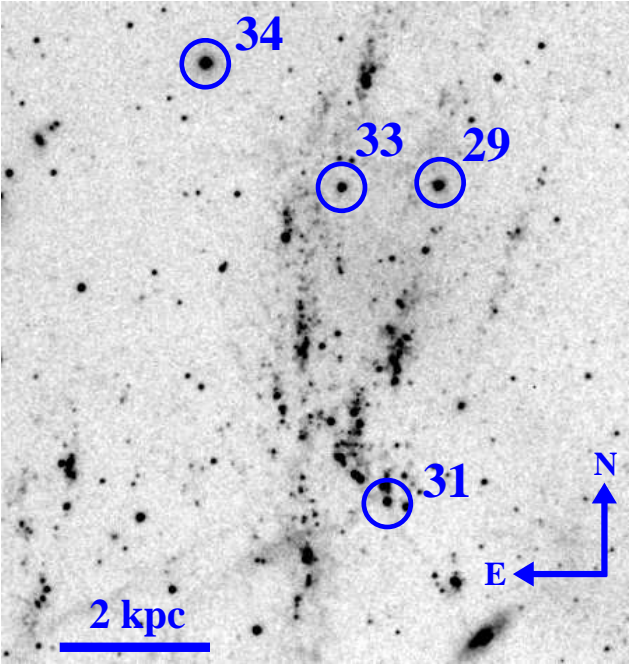}
s\caption{$R$ band image showing a star forming filament near NGC 1275. The background gradient due to NGC 1275 has been removed. UCDs  in this region are circled, with their identification numbers from Table~\ref{tab1275} also shown. A solid bar of length 2 kpc is provided for scale. Embedded in the filaments are numerous massive, young, blue star clusters. UCDs 29 and 31 are bluer than typical UCDs with $(B-R)_{0} < 0.6$.}
\label{colim}
\end{figure}

Several of the red UCDs in the vicinity of NGC 1275 are also located in the regions of the filaments. Red UCDs at the projected distance of these filaments must have survived at least 1 Gyr since their formation without being tidally disrupted or destroyed by their host galaxy. Fig.~\ref{colim} shows examples of blue and red UCDs residing in or near filaments, along with clumpy regions of star formation that might be responsible for the formation of at least some UCDs and GCs (see the discussion in Section 6). 

We consider the fact that we may be missing some young, blue star clusters due to patches of dust visible in NGC 1275, which would cause internal reddening. As a result of this dust, a possible source of contamination in our ``normal'' UCDs could arise from bright, young massive star clusters that have been reddened by this dust. We consider this scenario unlikely as we see no bright blue star clusters that are round in shape in regions far away enough from the galaxy core that such internal reddening is unlikely to occur. It can be seen from Fig.~\ref{both_cmds}  and Table~\ref{tab1275} that the youngest blue UCD around NGC 1275 has$(B-R)_{0} \sim$0.3. For a blue star cluster to be sufficiently reddened to move it onto the blue or red subpopulations requires a colour shift $\sim$1 mag. Such an object would also be dimmed by a similar amount. Assuming a similar correction to the Galactic reddening correction applied in the direction of Perseus ($A_{B}=0.435$, $A_{R}=0.703$), there are no blue objects around NGC 1275 sufficiently bright to masquerade as red UCDs reddened by their host galaxy. Similarly, even with a reddening vector $A_{V} = 1$, there are no objects today to be reddened into the colour range exhibited by ``normal'' UCDs (see Fig.~\ref{cols1275}). 

\subsection{UCDs in the rest of the cluster}

\begin{figure}
\includegraphics[width=0.48\textwidth]{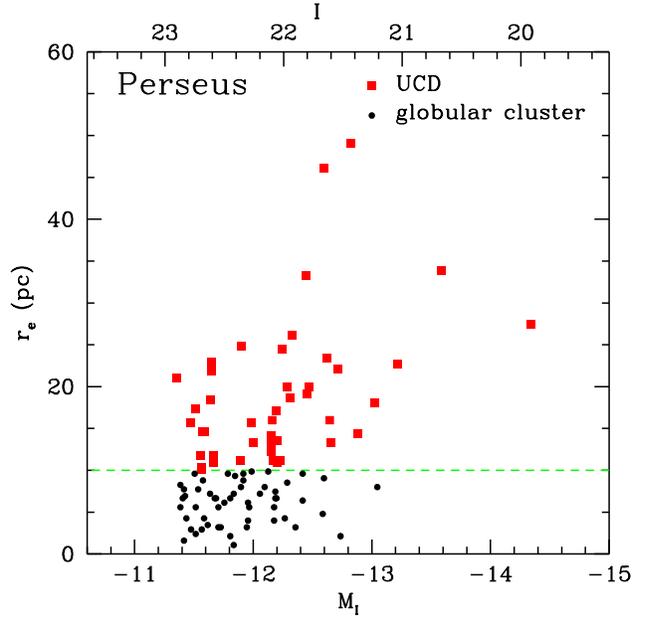}
\caption{Size-magnitude diagram for compact objects identified in the  regions covered by our $F555W$ and $F814W$ band $HST$ ACS imaging. The symbols and lines have the same meaning as for Fig.~\ref{sizes1275}}
\label{sizesper}
\end{figure}

\begin{figure}
\includegraphics[width=0.47\textwidth]{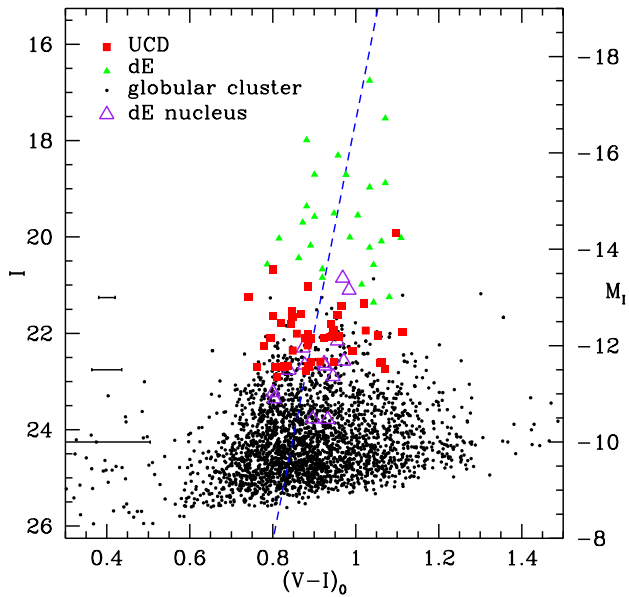}
\caption{Colour magnitude relation (CMR) for all UCDs identified from the $F555W$ and $F814W$ band imaging (red squares). Globular clusters ($r_{e} < 10$ pc) are shown as black dots. The Perseus dwarf elliptical sample identified in \citet{penny09} and \citet{derijcke09} is also included for comparison as green triangles.  Typical error bars at various magnitudes for the colours of the UCDs and GCs are given. The blue dashed line is a fit to the colour magnitude relation for dEs brighter than $M_{I} = -13.5$. The UCDs are consistent with the dE CMR extrapolated to lower magnitudes. dE nuclei are also shown as blue triangles. This plot is best viewed in colour in the online version of the journal.}
\label{UCDsvi}
\end{figure}

We also identify a total of 44 candidate UCDs in our F555W and F814W imaging, which includes the bright elliptical NGC 1272. As for the NGC 1275 region, we select UCD candidates to have sizes 10~pc~$< r_{e} < 150$~pc, and we make no selection based on either luminosity or colour. These objects have sizes 10~pc~$< r_{e} < 49$~pc, magnitudes brighter than $M_{V} = -11$, and colours $0.7 < (V-I) < 1.1$. Their mean ellipticity $b/a$ is 0.87. The sizes of these objects are shown in Fig.~\ref{sizesper}, with their colours and magnitudes included in Fig~\ref{UCDsvi}. Also included on this plot are the colours of dEs in the cluster core identified in \citet{derijcke09} and \citet{penny09}, along with GCs identified in this work. The colours and magnitudes of Perseus dE nuclei are also shown to investigate any similarities between UCDs and dE nuclei. To compare the colours of the UCDs and dEs, we perform a least squares fit to the dE population for all dEs brighter than $M_{I} = -13.5$. Dwarf ellipticals fainter than this are excluded due to a small sample size at the faint end. The UCDs follow the extrapolated colour magnitude  relation of the dEs down to $M_{I} = -11$.  

The colour magnitude relation for dwarf ellipticals is interpreted as a mass-metallicity relation, such that brighter dwarfs are more metal rich and therefore have redder colours. In \citet{penny08}, we found that the spectra of cluster dEs are consistent with ages $>2$ Gyr, and therefore have little age information left in their colours. The more luminous galaxies have deeper potentials and therefore retain their metal content more effectively. As a result, more massive dwarfs can self-enrich to higher levels than less massive ones.  UCDs continue this mass-metallicity relation to fainter magnitudes. dE nuclei and UCDs also occupy the same region of colour-magnitude space, as has been noted by other authors (e.g. \citealt{brodie11}).

\subsection{The UCD size magnitude relation}

\begin{figure*}
\includegraphics[width=0.48\textwidth]{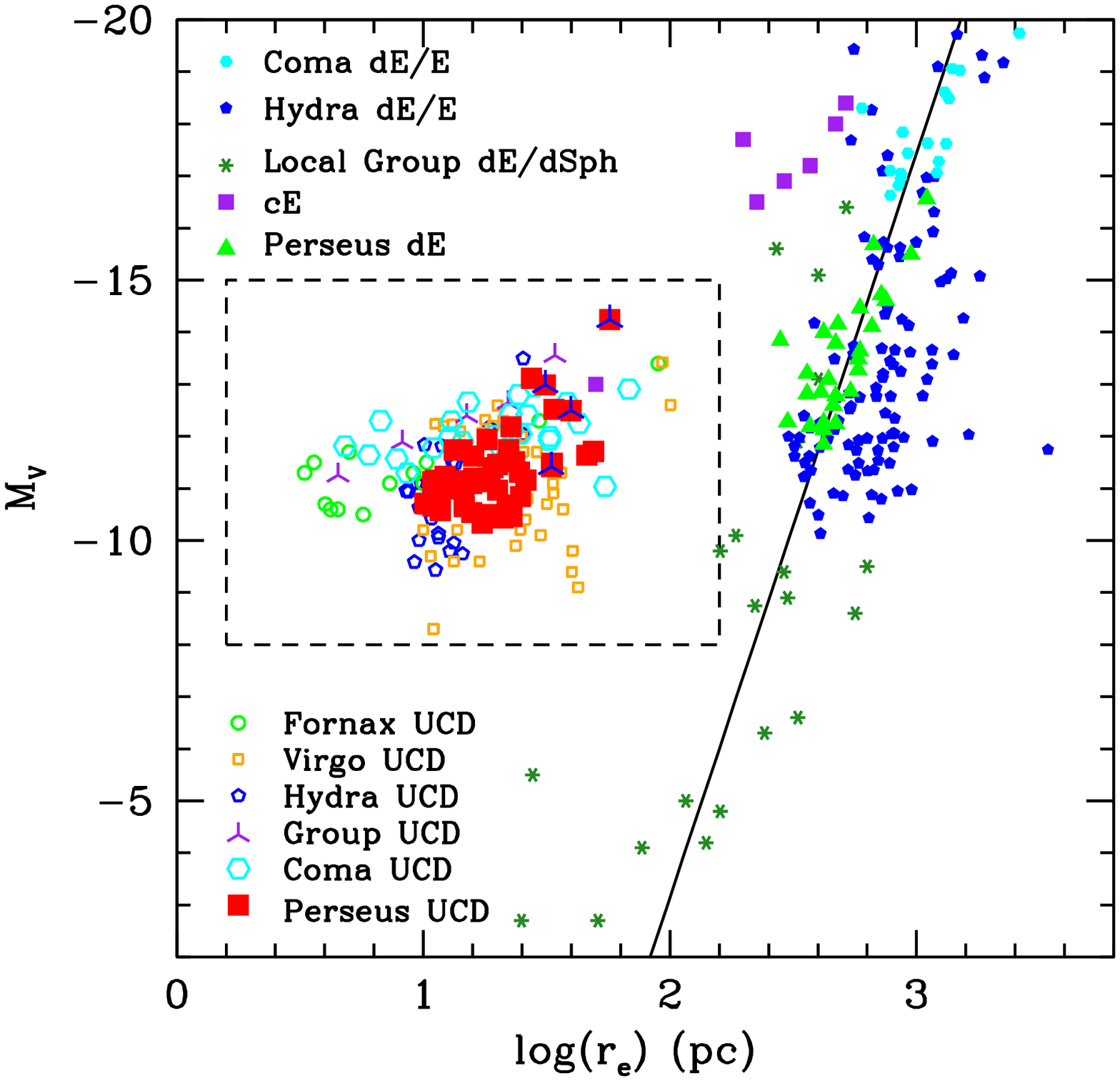} \includegraphics[width=0.48\textwidth]{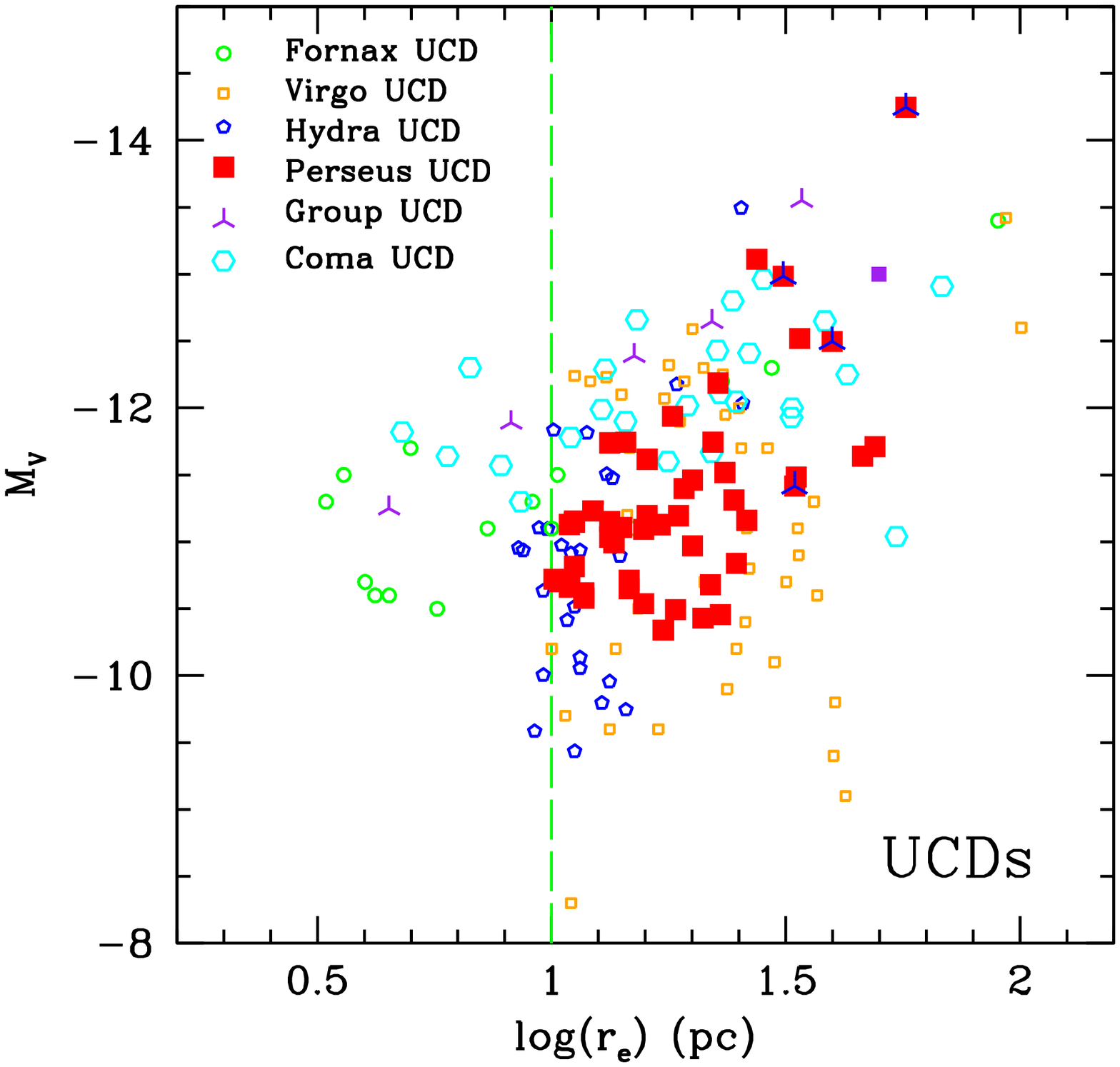}
\caption{Absolute luminosity vs effective radius for UCDs identified in our Perseus $F555W$ and $F814W$ band imaging. The Perseus UCDs are represented by red squares.  The red squares containing blue crosses are large UCDs ($r_{e} > 30$~pc) identified from the F435W and F625W imaging around NGC 1275 that have had their $B$ band luminosities converted to $V$ band using the conversion of \citet{fukugita95} for an elliptical galaxy. The green dashed line in the right hand plot corresponds to $r_{e} = 10$~pc. Additional data from the literature as described in the main text of the paper is also included. The left hand plot shows the sizes of UCDs along with cEs and dEs/dSphs, with the region of the plot occupied by UCDs enclosed by the dashed box. The right hand plot showing UCDs only (the region surrounded by a dashed box in the left hand plot). The solid line in the left hand plot is a linear fit to the size-luminosity relation for dwarf ellipticals is from \citet{derijcke09}. This plot is best viewed in colour on the online version of the paper.}
\label{sizecomp}
\end{figure*}

We compare the sizes and colours of the UCDs we have identified to those found for UCDs in other groups and clusters, along with the sizes found for dwarf ellipticals and compact galaxies taken from the literature in Fig.~\ref{sizecomp}. Much of our literature sample matches that of \citet{misgeld11a}. We add to this sample  our own data for the sizes of the Perseus dEs \citep{derijcke09}, and include the UCD sample from \citet{brodie11}.  Also included on this plot are the four UCDs identified from $R$ band imaging in the vicinity of NGC 1275 with sizes $>30$ pc. Their $B$ band magnitudes are converted to $V$ using the conversion of $B-V = 0.96$ from \citet{fukugita95} for an elliptical galaxy. 

The data we use is as follows: \citet{brodie11} and \citet{evstigneeva08} for Virgo UCDs,  \citet{mieske08} for Fornax UCDs, \citet{chiboucas11} for Coma UCDs, \citet{misgeld11a} for Hydra compact objects,  \citet{misgeld08} for Hydra dEs, \citet{norris11} for UCDs in nearby galaxy groups (around NGC 3923, NGC 4546 and the Sombrero Galaxy), \citet{graham03} for the Coma dEs,  \citet{martin08} for Local Group dSphs, \citet{gilmore07} for the sizes of Local Group objects, with \citet{grebel03} for the photometry of these objects. We also utilise sizes from \citet{derijcke09} for the Perseus dEs. All the UCD data sets included in Fig.~\ref{sizecomp} have spectroscopically confirmed cluster membership. 

We convert the $B$ band colours of \citet{graham03} to $V$ band using the $B-V$ colour magnitude relation constructed as used in \citet{derijcke09}. This relation changes from $B-V \sim 0.7$ at $M_{B} = -8$ to $B-V \sim 0.8$ at $M_{B} = -22$. We therefore apply a constant mean colour correction $< B - V > = 0.8$ for this data, as most of it is at the bright end of the CMR.    We perform a similar conversion for the UCD sample of \citet{madrid10}.

The sizes and magnitudes for the UCDs we identify in this study agree well with those found in other regions of the Universe, which all have spectroscopically confirmed group/cluster membership. We find no low luminosity, extended GCs/UCDs that might fill in the gap between the sequence of galaxies and compact star clusters i.e. objects with $r_{e} \sim 100$~pc and $M_{V} < -15$. Objects with more extended effective radii and faint magnitudes would have very low surface brightness, making them difficult to detect in existing surveys of higher redshift ($D >30 $ Mpc) clusters such as Perseus. In addition, the gap of objects with sizes 80~pc~$< r_{e} < 150$~pc remains on the sequence formed by cEs and UCDs.

\subsection{The distribution of UCDs in the Perseus Cluster}

\begin{figure}
\includegraphics[width=0.45\textwidth]{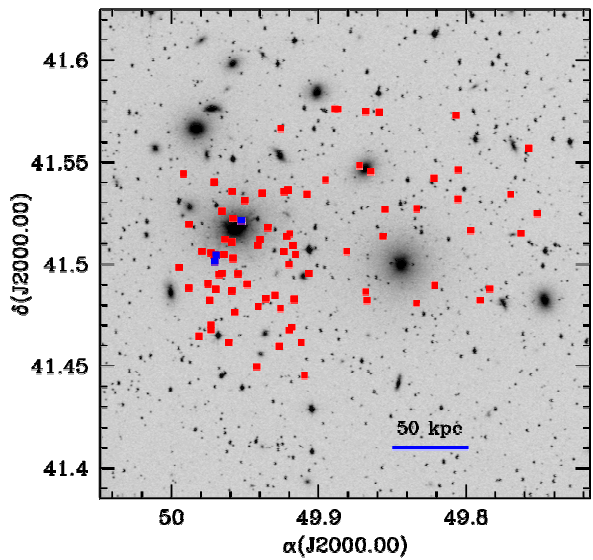}
\caption{The positions of all compact stellar systems in the Perseus Cluster with sizes 10~pc$ < r_{e} < 150$~pc that make our final catalogue of UCDs. The red squares are UCDs with ``typical'' colours, and the blue boxes are UCDs with very blue colours consistent with a young stellar population. The solid bar is 50 kpc in size. UCD candidates are predominantly found in the very centre of the cluster around NGC 1275. The imaging is not continuous, so further UCDs likely exist in the cluster core.} 
\label{UCDdis}
\end{figure}

The locations of all UCDs in the Perseus Cluster covered in our imaging are shown in Fig.~\ref{UCDdis}. It can been seen that UCDs are preferentially found in the highest density regions of the cluster. In particular, the region of the cluster around NGC 1275 appears to have a very large population of UCDs. This large population implies the formation of UCDs in Perseus may be linked to NGC 1275, or that UCDs lie in the deepest part of the Perseus potential well. Thus massive star clusters may be forming in the filaments surrounding the galaxy, as suggested by the close projection of NGC 1275 UCDs and the filaments. Alternatively, these objects could be formed via the tidal stripping of nucleated dwarf ellipticals in the very densest regions of the cluster. Formation scenarios for the Perseus UCD population are discussed further in $\S$~\ref{sec:discussion}.

\section{Comparing the Perseus UCD and GC populations}
\label{sec:gcs}

UCDs have been hypothesised to be an extension of the globular cluster population to larger sizes and brighter magnitudes.  If they originate from the same population, then GCs and UCDs might be expected to share similar properties, e.g. exhibiting similar colour-distance, radial surface density and colour-magnitude trends. To search for similarities between the two populations, we therefore compare the UCDs we have identified to the Perseus GC system. 

The cores of rich clusters contain some of the most massive galaxies in the Universe, which have large globular cluster systems. Such GC systems have bimodal distributions, with blue and red subpopulations (e.g. Fig.~\ref{NGC1275_fit}). The blue subpopulation may also reveal a blue tilt, i.e. redder colours at brighter magnitudes. This ``blue tilt'' is likely a mass-metallicity relation, with the least massive GCs being the most metal poor. It is thought that more massive GCs on this sequence are able to self-enrich to a small extent, thus increasing their metallicity and resulting in the observed tilt \citep{strader08}. The red subpopulation typically exhibits no such tilt. We investigate if these subpopulations can be identified in Perseus and if UCDs follow these colour trends. 

We define GCs to be all objects in our cleaned catalogues with sizes $1 < r_{e} < 10$ pc and $S/N > 5$. The lower size limit ensures that stars are not included in our GC sample, with the upper limit of 10 pc to separate of UCDs and GCs. These GCs are drawn from the same visually cleaned catalogues as our UCD sample. To ensure we are sampling GCs rather than background objects, we limit our GC sample to those around massive galaxies where the probability of identifying GCs rather than compact background galaxies is higher.  Perseus contains two massive galaxies in its core for which we have (reasonably) complete coverage: NGC 1275 and NGC 1272, therefore we examine the GC systems of these two massive galaxies.

\subsection{The globular cluster system of NGC 1275}
\label{sec:ngc1275}

The initial colour magnitude diagram prior to cleaning for all objects detected in the field of NGC 1275 reveals two distinct locii, with a sequence of blue objects not seen in other regions of the cluster.  The latter are associated with star forming regions in the filaments of NGC 1275. This has been observed by several authors to date, e.g. \citet{canning10}. After manual cleaning of clumpy objects in this field, the blue objects mostly disappeared. The remaining objects exhibit typical globular cluster colours ($0.9 < (B-R)_{0} < 1.7$).

\begin{figure}
\includegraphics[width=0.45\textwidth]{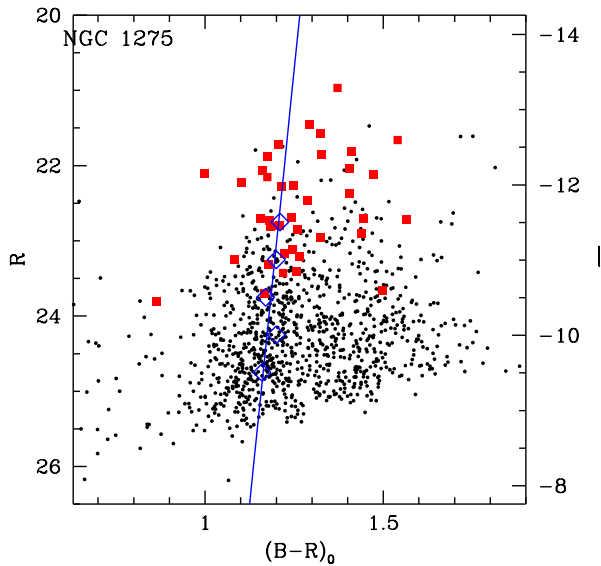}
\caption{The colour magnitude diagram for GCs within a projected distance of 50 kpc from NGC 1275. The blue line is a weighted least squares fits to the blue GC subpopulations. The large open diamonds represent the mean colours of the blue subpopulation as determined using \textsc{rmix}. A significant blue tilt is seen. The fits to the red subpopulation is not included as we do not find a statistically significant red tilt. UCDs in this region are plotted as red squares, and the majority can be seen to lie on the blue and red GC subpopulations.}
\label{NGC1275_fit}
\end{figure}

The remaining objects do not have a simple colour distribution. It can be seen from Fig.~\ref{cols1275} that the globular cluster system of NGC 1275 appears to contain red and blue subpopulations. To investigate these subpopulations, we examine the colour distribution for globular clusters around the galaxy.  

We separate the GC population of NGC 1275 into its blue and red subpopulations to determine the blue and red tilts for this galaxy. The GCs are first binned by magnitude into bins of width 0.5 mag. The blue and red peaks of each magnitude bin are then determined using \textsc{rmix}, a  package in the programming language \textsc{r}, using two Gaussian distributions with fixed $\sigma$. We then perform a fit to the colour peaks of the magnitude bins. The very brightest compact objects with $R > 22$  are excluded from this fit due to the small number of objects in this magnitude bin. 

The fit to the blue GC subpopulation is found to be:

\begin{equation}
(B-R)_0 = (-0.022 \pm 0.008)R + (1.70 \pm 0.196)
\end{equation}

\noindent The fit to the GC red subpopulation:
\begin{equation}
(B-R)_0 = (0.015 \pm 0.011)R + (1.10 \pm 0.264)
\end{equation}

The result of the fit to the blue subpopulation is included in Fig~\ref{NGC1275_fit}. A blue tilt is seen for the blue subpopulation, such that fainter objects are bluer than brighter ones. Within the errors, the fit to the red subpopulation has zero slope, suggesting no mass metallicity relation for these objects.

We also include the UCDs in Fig~\ref{NGC1275_fit}. The UCDs occupy the bright-end of the GC colour-magnitude trend, and lie on the blue and red subpopulations traced by the galaxy's GC system. These colour-magnitude trends hint that Perseus UCDs might be simple massive GCs.

\subsubsection{Radial colour trends in the NGC 1275 GC system}

\begin{figure}
\includegraphics[width=0.45\textwidth]{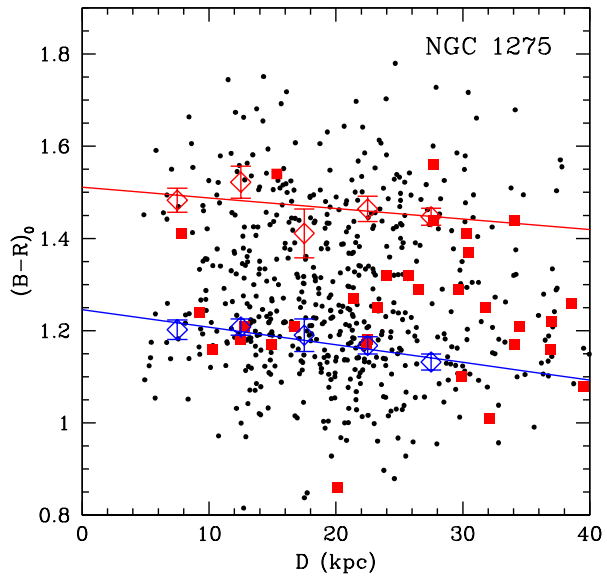}
\caption{$(B-R)_{0}$ GC colour versus projected distance from NGC 1275. Blue objects with colours $(B-R)_{0} < 0.8$ (i.e. star forming) are excluded from this plot for clarity. The data are binned by galactocentric distance, and the mean colours of the blue and red subpopulations are determined using the \textsc{r} package \textsc{rmix}. The solid lines are linear fits to the binned data. Clear radial trends in colour are seen, with both the blue and red subpopulations becoming bluer with increasing distance from the galaxy centre, suggesting a radial metallicity gradient. The UCDs are included as red squares, with a fraction of these objects following the GC colour-distance relationships. }
\label{radcols1275}
\end{figure}

We examine the radial trend in the colour of the NGC 1275 globular cluster system in Fig.~\ref{radcols1275}. First, the projected distances of each globular cluster from NGC 1275 is calculated, and a plot of colour versus projected distance from NGC 1275 is made (Fig.~\ref{radcols1275}). The relations between colour and projected distance for the two subpopulations are fit separately. The GC subpopulations are binned by galactocentric distance, with bins of 5 kpc, and the red and blue means of these bins are then found using the same method as for the blue and red tilt fitting. The peaks of each of the radial distance bins are then fit with a weighted least squares fit, with each point weighted by $1/{\sigma}^{2}$ . Objects at a small projected distance from NGC 1275 are excluded from this fit due to the small number of globular clusters identified in this region. 

The radial colour trend in the blue subpopulation is:

\begin{equation}
(B-R)_{0} = (-0.0038 \pm 0.0012)D + (1.25 \pm 0.024)
\end{equation}

\noindent and for the red subpopulation:

\begin{equation}
(B-R)_{0} = (-0.0023 \pm 0.0015)D + (1.51 \pm 0.032)
\end{equation}

\noindent where $D$ is the projected distance from NGC 1275. Within the errors, there is no significant difference in the colour gradients of the two subpopulations as has been found in previous studies \citep{forbes11}.  

These radial colour trends are likely due to metallicity gradients.  The globular clusters are expected to trace the metallicity gradient of their host halo, as globular clusters typically host old stellar populations.

The colours and projected distances of the UCDs from NGC 1275 are also included as red squares in Fig.~\ref{radcols1275}.  Approximately half of the UCDs follow the colour-distance relation traced by the blue and red GC subpopulations. However, many UCDs at $D>25$~kpc have colours intermediate between the two GC subpopulations with $(B-R)_{0}\sim 1.3$. A least squares fit to the UCD colour-distance relation for all UCDs redder than $(B-R)_{0} = 0.8$ has a slope of $0.0012 \pm 0.019$, consistent with zero. Thus we find no radial colour trend in the UCD population around NGC 1275.

\subsection{The globular cluster system of NGC 1272}

The giant elliptical NGC 1272, located at $\alpha$ = 03:19:21.3, $\delta$ = +41:29:26, with $M_{B} = -22.45$ \citep{refcat} is the second brightest galaxy in the Perseus Cluster. This galaxy is a normal elliptical, and unlike NGC 1275, is relatively easy to fit and remove from our imaging using \textsc{iraf ellipse}. This allows for the identification of compact stellar systems to smaller galactocentric distances.

\begin{figure}
\includegraphics[width=0.48\textwidth]{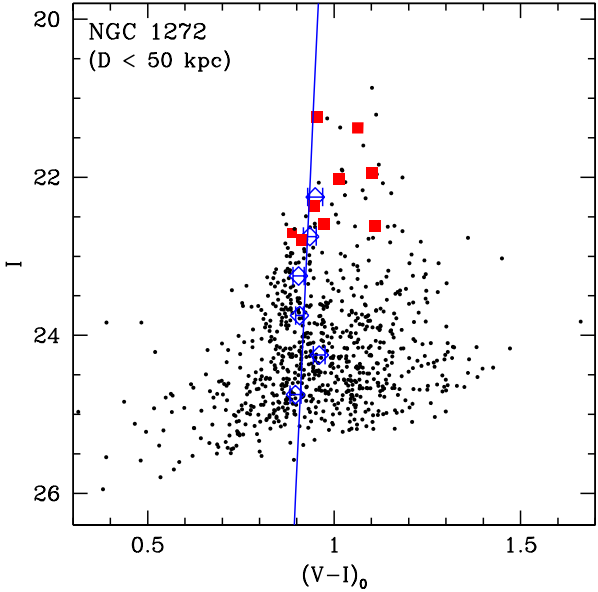}
\caption{The $(V-I)_{0}$ colour magnitude relation for the GC and UCD populations of the second brightest cluster galaxy NGC 1272, out to a distance of 50 kpc. At brighter magnitudes, two colour sequences are seen, corresponding to the red and blue GC subpopulations. The solid line is a weighted least squares fits to the blue GC subpopulation as determined via Gaussian fitting using \textsc{rmix}. As for NGC 1275, a blue tilt in the globular cluster population is seen. No statistically  significant red tilt was found.}
\label{cmdgc1272}
\end{figure}

We fit the blue and red GC subpopulations using the same method as for NGC 1275, with the resulting fit to the blue tilt shown in the colour magnitude diagram for NGC 1272 GCs (Fig~\ref{cmdgc1272}). 

The fit to the blue GC subpopulation is:

\begin{equation}
(V-I)_0 = (-0.098 \pm 0.008)I + ( 1.15 \pm 0.192)
\end{equation}

\noindent For the red GC subpopulation:

\begin{equation}
(V-I)_0 = (0.011 \pm 0.013)I + ( 0.88 \pm 0.299)
\end{equation}

The GC system of NGC 1272 also exhibits red and blue subpopulations, with the blue subpopulation exhibiting a blue tilt, with the slope of the fit to the red subpopulation exhibiting a slope consistent with zero within the errors.  Similar to NGC 1275, the UCDs in the vicinity of NGC 1272 occupy the bright end of the colour magnitude relation, falling on both the blue and red subpopulations. This further strengthens the argument for UCDs being an extension of the GC population to larger sizes. 

As for NGC 1275, we investigate radial colour trends in the globular cluster colours by examining the colours of the galaxy's globular cluster system versus distance from the galaxy centre. We examine the globular cluster system of this galaxy out to $R = 50$ kpc, which is the approximate midpoint of the projected distance between NGC 1272 and its nearest neighbouring massive elliptical to avoid confusion between the two galaxies globular cluster systems. Unlike NGC 1275, we cannot distinguish the two subpopulations in this radial colour distribution, with a fit to the GC system as a whole given by:

\begin{equation}
(V-I)_{0} = (-0.0029 \pm 0.0005)D + (0.988 \pm 0.0117)
\end{equation}

\noindent where $D$ is the distance from the galaxy centre in kpc.  

The colours of the galaxy's globular cluster system become increasingly bluer with galactocentric distance. However, this may simply reflect the changing mix of GC subpopulations with galactocentric distance, such that the blue subpopulation extends to larger radii than the red subpopulation.

\section{Discussion}
\label{sec:discussion}

We have identified a large population of 84 UCD candidates out to 250~kpc distance from the Perseus Cluster centred galaxy NGC 1275.  The colours of these UCDs are, in general, consistent with old, passive stellar populations, lying on an extrapolation of the dE colour magnitude relation to fainter magnitudes. Furthermore, Perseus dE nuclei and UCDs have very similar colours, in agreement with the results of \citet{brodie11}. This result suggests a possible evolutionary link between UCDs and dE nuclei. However, UCDs also show similarities to GCs in terms of their colour distribution, such that they lie on the bright ends of the blue and red GC subpopulations. 

In terms of size-luminosity trends, we confirm the findings of Brodie et al. (2011) that UCDs occupy a broad locus in size-luminosity parameter space. A size-luminosity trend is seen for the Perseus UCDs, in that the largest UCDs have the highest luminosities, but at fainter magnitudes ($M_{V}$ fainter than $-12$), the spread in the distribution of UCD sizes at a given magnitude is large, such that these objects do not show a well-defined size-luminosity relation. 

We identify in our sample a number of ``extended'' UCDs with sizes between 30~pc and 57~pc. \citet{gilmore07} suggest that star clusters have characteristic sizes $<30$~pc, whereas all larger galaxies are galaxies with sizes $>120$~pc, with no objects having sizes 30~pc $< r_{e} < 120$~pc.  This result is not in agreement with the sizes of the more extended UCDs we identify in Perseus.  Our size measurements are complimented by those of \citet{brodie11} who find examples of such objects in Virgo with spectroscopically confirmed redshifts, further blurring the boundary between massive star clusters and compact galaxies. Although the nature of UCDs remains unknown, they are filling the scale-size gap between galaxies and star clusters proposed in \citet{gilmore07}.

One possible formation scenario for the Perseus UCD population is that they are an extension of the GC system to brighter magnitudes. This is discussed in more detail below. We also discuss two alternative formation scenarios for these objects: they are the remnants of nucleated dwarf ellipticals that have been tidally stripped by massive galaxies and the cluster potential; or they are formed in the star forming regions of the filaments surrounding NGC 1275. 

\subsection{UCDs as massive globular clusters}

The Perseus UCD and GC populations both follow the same radial surface distributions at distances $D > 30$~kpc from NGC 1275 (Fig.~\ref{rdist}), suggesting they are part of the same distribution. At smaller galactocentric distances, the radial surface density of UCDs flattens off, suggesting these objects are either disrupted at small distances, or cannot form in the increased tidal potential of their host galaxy. 

The UCDs in the vicinity of NGC 1275 and NGC 1272 follow the same colour-magnitude trend as the GC systems of these galaxies, such that there are blue and red subpopulations of Perseus UCDs. Around NGC 1275, there is also evidence that a fraction of UCDs follow the colour-distance trend of the GC population, such that bluer objects are found at larger projected radii from the massive galaxy. These similarities suggest that some UCDs in Perseus might be massive globular clusters. 

We find UCDs to be consistent with the colour-magnitude trend of dEs to fainter magnitudes and blue GCs to higher magnitudes. So unlike Brodie et al. (2011) we did not find evidence for the UCDs to be offset (to bluer colours at a given magnitude) compared to the blue GCs. However we note that our overlapping magnitudes for both UCDs and GCs is small.

\subsection{UCDs as tidally stripped of nuclei dwarf ellipticals}

The colours of Perseus UCDs are also remarkably similar to those of dE nuclei, similar to the findings of \citet{brodie11}. One formation mechanism for UCDs is through the tidal stripping of nucleated dwarf ellipticals \citep{bekki01,drinkwater03}. UCDs have been observed to have a range of $\alpha$ enhancements \citep{brodie11}. This is also consistent with dEs, which are known not to be a simple, homogeneous population, but rather exhibit a range of ages and metallicities inconsistent with a single formation scenario \citep{penny08}. We therefore consider the possibility that dE nuclei are the progenitors of UCDs by considering the Perseus dE population.

The faint-end slope of the luminosity function in Perseus is remarkably shallow at $\alpha = -1.26$ \citep{penny08}, suggesting the core of Perseus may be dwarf depleted. The destruction of dwarf ellipticals via tidal processes is once such way to explain this shallow faint-end slope. Nucleated dwarf ellipticals are abundant in Perseus, and such galaxies have been suggested as the progenitors of UCDs \citep{bekki01,drinkwater03}. Furthermore, cD/BCG galaxies likely reach their large masses via numerous mergers and tidal interactions with other cluster members (e.g. \citealt{brough11}). The low density outer regions of nucleated dwarf ellipticals would be easily stripped, with the more compact nuclei left behind. Very deep imaging to extremely low surface brightness ($\mu_{V} = 29$ mag arcsec$^{-2}$) is required to search for the tidal debris that would result from the stripping of the stellar light envelopes of nucleated dwarf ellipticals. 

Using the technique of \citet{penny09}, we can estimate how close a typical nucleated dwarf elliptical must come to the cluster cenre for stars at it's visible (Petrosian) radius to be stripped by the cluster potential. By estimating the strength of the tidal potential acting on a star located at the Petrosian radius of the dwarf, we can estimate how massive the dwarf must be to prevent its disruption by the tidal potential (or host galaxy). We calculate this for a typical dwarf elliptical of size $r_{e} = 1.5$ kpc and mass $10^{8}$ M$_{\odot}$. This dwarf is dE 10 from \citet{cgw03}, and is located at a projected distance of 193~kpc from NGC 1275.

\begin{figure}
\includegraphics[width=0.47\textwidth]{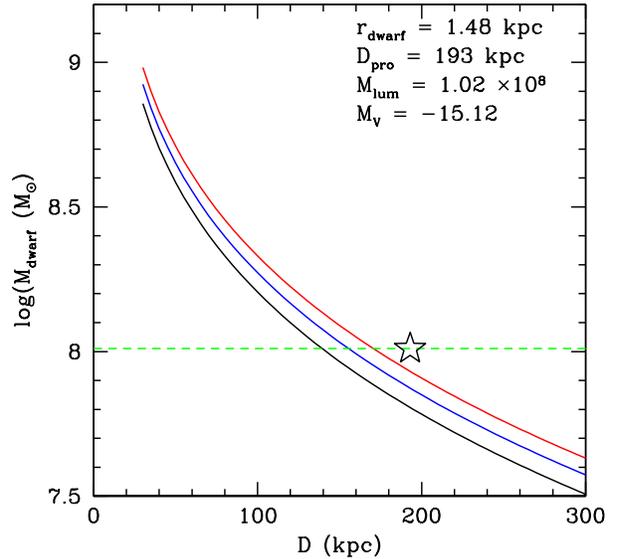}
\caption{The minimum stellar mass a nucleated dwarf elliptical must have to prevent stars at the Petrosian radius of the dwarf being stripped by the cluster potential as a function of cluster-centric distance. The three lines represent (from bottom to top) orbital eccentricities of 0, 0.5 and 1.  These masses were calculated using the method of \citet{penny09} for a Perseus nucleated dwarf elliptical of size 1.5~kpc, located at $03:19:00, +41:29:02$ (dE 10 in \citealt{cgw03}).  The stellar mass of the dE in this example is $10^{8}$ M$_{\odot}$. The green dashed line is the stellar mass of the dwarf, and the star is located at its current projected distance from the cluster centre (193 kpc). At small distances from the cluster centre ($<140$ kpc), a  typical nucleated dwarf elliptical will be easily stripped of its outer stars unless it has a large enough dark matter mass to prevent this.}
\label{totmass}
\end{figure}

Fig.~\ref{totmass} shows that at distances $<100$ kpc from the cluster centre, a typical nucleated dE will not have enough visible mass to retain stars in its outer region. Therefore, it is plausible that a number of the UCDs we see in clusters today are the remnant nuclei of dwarf ellipticals that did not reside in sufficiently massive dark matter haloes to prevent their tidal disruption. Nucleated dwarf ellipticals would therefore be easily stripped of their outer stellar envelopes via tidal processes, with only their dense nuclei, that have sizes similar to UCDs, remaining to the present day.  The fact that number density of UCDs is flat in the inner regions of NGC 1275 (Fig.~\ref{rdist}) does not necessarily rule out a galaxy stripping origin for UCDs. If tidal forces are sufficiently strong in the inner regions of the galaxy, more extended clusters or nuclei will be stripped to smaller radii by tidal effects.   

The H$\alpha$ filaments present in the deeper $R$ band imaging make searching for tidal structures that would indicate tidal interactions at small galacto-centric distances difficult.  The $B$ band imaging is free of these filamentary structures, and exhibits faint shells, indicating that galaxy-galaxy interactions have taken place here. NGC 1275 is a shell galaxy \citep{holtzman92,conselice1275}, with these shells formed via the accretion of satellite galaxies which contribute to the mass build up of the BCG.

\begin{figure}
\includegraphics[width=0.47\textwidth]{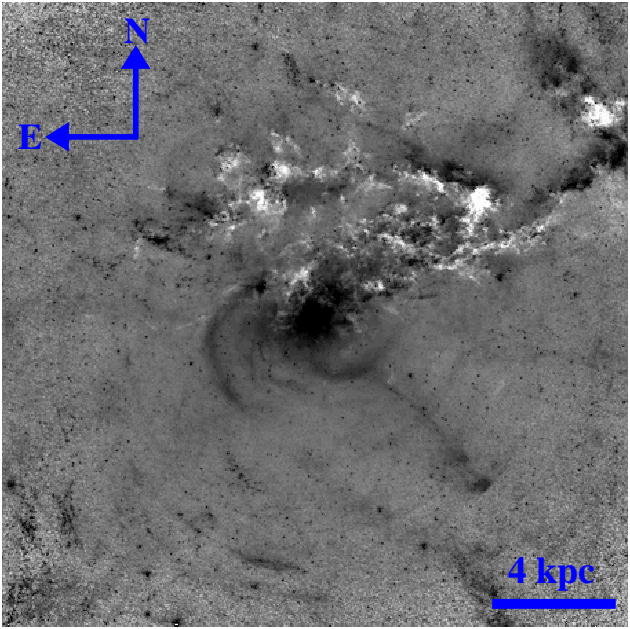}
\caption{The shells around NGC 1275 highlighting low surface brightness features in the galaxy's halo. This image was created by dividing the original image by a smoothed version of itself to highlight fine structures. Several shells can be seen to the top right hand side of the galaxy, with a long plume-like structure intersecting the shells. These shells indicate that past galaxy-galaxy interactions have taken place in this region of the cluster core. The patchy dust to the North of the image belongs to the foreground high velocity system. }
\label{shells}
\end{figure}

To highlight these shells, we perform an unsharp mask subtraction on this imaging. Features are highlighted by dividing  the original image by a smoothed version of itself. The image was filtered using a $300 \times 300$ pixel median filter using \textsc{iraf imfilter}. The original image was then divided by the filtered image to remove large scale structure (such as galaxy light), with smaller scale features such as shells and dust lanes remaining. The result of this subtraction is shown in Fig~\ref{shells}. Numerous low surface brightness shells are visible in this imaging, with the brightest shells located towards the centre of the galaxy, with fainter shells visible at larger radii. 

Such interactions would disrupt the diffuse stellar envelopes of dwarf ellipticals more easily than their dense nuclei. While the remnants from such interactions would be impossible to detect in our existing imaging due to their extremely low surface brightness, nevertheless tidal interactions likely play an important role in the cluster core. 

The majority of nucleated dwarfs in Perseus are very smooth, round systems, with no evidence for asymmetry in their structures that would indicate interactions with the cluster tidal potential \citep{penny09}. However, these may just be the dEs that are dark matter dominated enough to survive unperturbed to the present day. A fraction of the UCD population of Perseus may therefore be the remnants of stripped dEs.

\subsection{UCD formation in filaments}

\begin{figure}
\includegraphics[width=0.47\textwidth]{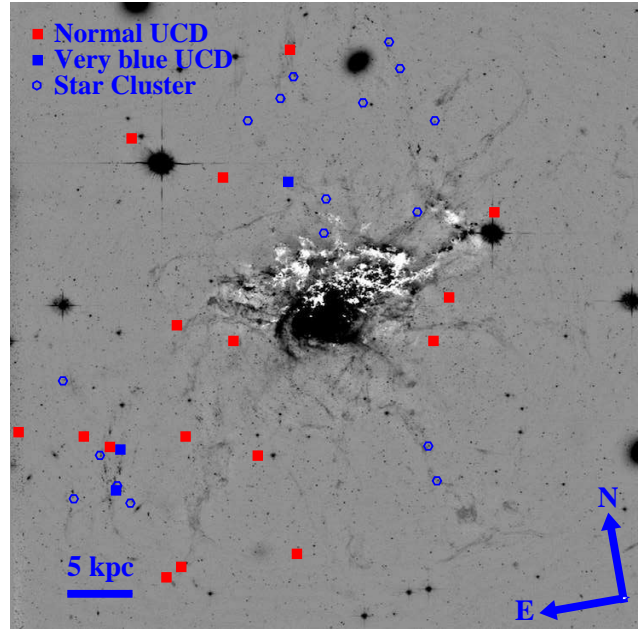}
\caption{An unsharp mask subtracted image of the F625W imaging of NGC 1275. This image highlights the complex system of filaments that surround the galaxy. Overplotted on the image are the positions of the UCDs within this field of view. The blue filled squares are those UCDs that have colours $(B-R)_0 < 0.6$. The red filled squares are UCDs with typical colours. The blue open hexagons are objects with round morphologies reminiscent of GCs/UCDs and $S/N > 40$, but without reliable size measurement as they reside in crowded regions.}
\label{filaments}
\end{figure}

Given the proximity of NGC 1275 UCDs to the gaseous filaments surrounding this galaxy, we investigate these filaments as a possible site for UCD formation.  NGC 1275 is an unusual galaxy, and likely represents a good analogue for processes that occurred at higher $z$ for most brightest cluster/group galaxies, as it is still actively undergoing formation and evolution. NGC 1275 exhibits a highly complex structure (Fig.~\ref{filaments}), surrounded by a complex system of H$\alpha$ filaments. 

A number of the filaments surrounding NGC 1275  are associated with active star formation. Such star forming regions are likely formed when cooling gas collapses into thinner filaments, leading to star formation (McDonald et al. 2010). A similar region of star formation to those observed around NGC 1275 has been observed in Abell 1795, but otherwise remain essentially unstudied. This suggests that such filamentary star forming structures are relatively short lived, with supernova feedback rapidly removing the material for further star formation in these structures.  If such star forming filamentary structure were more common at higher $z$, this would naturally explain the old ages seen for the majority of UCDs.

The number of UCD candidates we identify in the vicinity of NGC 1275  could therefore be due to their formation in the cluster centred galaxy.  Massive compact blue star clusters are known to exist in the core of NGC 1275, which have been hypothesised to be protoglobular clusters with masses $10^{7}$ - $10^{8}$ M$_{\odot}$ \citep{brodie98}. However, star formation is not limited to the core of the BCG, but rather extends to a number of knots and loops of star forming regions that enshroud the galaxy. The mechanisms for star formation in these regions has already been discussed in detail elsewhere (e.g. \citealt{canning10}) so will not be examined further in this work. Given that star formation is actively ongoing in this massive galaxy, this may provide a source of new UCDs in the cluster.  

The star forming filaments contain numerous young, massive star clusters. \citet{canning10} estimate typical masses for the brightest star clusters of $10^{7}$ M$_{\odot}$, with the typical UCD masses being $10^{6}$ to $10^{8}$ M$_{\odot}$ \citep{mieske08}. Through mergers, these young, massive star clusters build up enough mass to obtain sizes and luminosities consistent with those of the most massive UCDs. The very blue UCDs are likely early on in their formation, before their stellar populations have aged enough to resemble those of normal UCDs identified in previous studies. 

The 3 very blue UCDs we identify are both located in the vicinity of NGC 1275, with ``normal'' UCDs distributed more evenly through the cluster (Fig~\ref{UCDdis}). Given the young ages of their stellar populations ($<0.1$ Gyr), they must have formed very recently, and will therefore will not have had enough time to migrate away from their formation site.  This strongly suggests that the young, proto-UCDs are linked to the ongoing star formation in the filaments.  Perturbations by other cluster galaxies provide a mechanism to distribute these objects through the cluster. 

A sequence of young, very blue objects with $(B-R)_{0} <0.5$ are seen in the colour magnitude diagram around NGC 1275 (Fig.~\ref{both_cmds}). A number of these blue star clusters are in regions that are too crowded for accurate size determination, along with more isolated young, blue GC-sized objects.  Further examples of proto-UCDs therefore exist in the vicinity of NGC 1275. The locations of these very blue, round objects, along with UCDs in the immediate vicinity of NGC 1275 are shown on an $R$-band unsharp mask image in Fig. ~\ref{filaments}.  These objects are all located on or near filaments, hinting these are their formation sites. 

The typical colour of a UCD is $(B-R)_{0} \sim1.3$, which corresponds to an age of $\sim$1~Gyr for its stellar population. \citet{canning10} argue that at least some of the star clusters in these regions of star formation are being tidally disrupted by their host galaxy due to their elongated structures. If this is the case, a low mass star cluster might not withstand the tidal forces of the host galaxy long enough for its colours to transform to those typical for a UCD. Through mergers,  young, blue star clusters could reach a sufficient mass such that they are able to survive the tidal forces of NGC 1275 to the present day to appear as UCDs in the clusters.  However, individual star clusters within the filaments are remarkably round, with ellipticity $\sim1$, showing they are already of sufficient mass to prevent tidal stripping and elongation by their host potential. A high mass, compact structure would be robust against the tidal forces in the cluster core, unlike the diffuse envelopes of dEs. Those with $(B-R)_{0} \sim$0.5 must already have survived 100 Myr without disruption, therefore the more massive UCD/GC sized objects in these filaments are resilient against such tidal interactions. 

To investigate this UCD formation mechanism, it will be necessary to investigate other clusters or groups that exhibit filamentary structure in the nearby (100~Mpc) Universe. If young, blue UCDs are seen, this would strengthen the hypothesis that such objects are formed in these filaments. These young, very blue UCDs furthermore highlight the importance of utilising a complete range of colours ranging from star forming to passive when carrying out searches for such objects.

\section{Conclusions}
\label{sec:conclude}

We identify in the Perseus Cluster core 84 UCD candidates. The majority of these objects have sizes and colours that are comparable with those of UCDs identified in previous studies. Three UCD-sized objects are bluer than typical UCDs, exhibiting $(B-R)_{0}$ colours of 0.28, 0.54 and 0.59. These colours are consistent with objects that have ceased star formation in the last 100~Myr, and are therefore still located close to their formation site. The very blue UCDs are at projected distances $<$25 kpc from the centre of NGC 1275, and are furthermore located near the star forming filaments that surround the galaxy. Numerous other large, round star clusters in star forming filaments too crowded for accurate size determination are also seen. 

Based on these objects, and the fact that UCDs in Perseus are remarkably centrally concentrated around the BCG NGC 1275, we hypothesise that a fraction of the UCD population are formed in the star forming filaments surrounding this galaxy.  Through mergers, massive star clusters in these regions are able to reach the masses and sizes typically seen for UCDs. Many star clusters within these filaments have colours of $(B-R)_{0}~\sim~0.5$, consistent with stellar populations of age $\sim100$~Myr.  These star clusters therefore must be robust against tidal perturbations to survive for sufficient timescales that their stellar populations become red. Over time, such objects will become more evenly distributed through the cluster due to perturbations from other cluster galaxies.  

We do not rule out other UCD formation mechanisms such as tidal stripping or massive GCs, based on similarities between UCDs and both dE nuclei and GCs. UCDs and GCs share a similar radial density profile, implying they are part of the same population. Numerous faint tidal features and shells are visible around NGC 1275, providing evidence that at least some of the UCDs in Perseus were formed by such tidal interactions.  

We compared the colours of the Perseus UCDs to globular cluster systems of NGC 1275 and NGC 1272 to investigate if UCDs are massive GCs. UCDs also lie on the bright end of the GC colour magnitude trend, following the blue and red subpopulations, suggesting they may be a bright extension of the GCs. It is therefore likely that a number of formation scenarios are required to explain the origin of UCDs in the cluster environment, strengthening the argument that UCDs are not a simple class of objects (c.f. \citealt{norris11}). We therefore conclude that although the Perseus cluster is an ideal environment to strip nucleated dE and hence form UCDs, we can not easily discriminate that origin from a massive star cluster one. Neither can we rule out a new formation path within gaseous filaments. 

\section*{Acknowledgments}

We thank the anonymous referee for their comments that helped improve this paper. SJP acknowledges the support of a DEEWR Endeavour Postdoctoral Award and an ARC Super Science Fellowship. CJC  acknowledges support from the STFC and the Leverhulme Trust. This research has made use of the NASA/IPAC Extragalactic Data base (NED) which is operated by the Jet Propulsion Laboratory, California Institute of Technology, under contract with the National Aeronautics and Space Administration. Some of the data presented in this paper were obtained from the Multimission Archive at the Space Telescope Science Institute (MAST). STScI is operated by the Association of Universities for Research in Astronomy, Inc., under NASA contract NAS5-26555. Support for MAST for non-HST data is provided by the NASA Office of Space Science via grant NNX09AF08G and by other grants and contracts.


\begin{thebibliography}{}

\bibitem[\protect\citeauthoryear{{Bekki}, {Couch} \& {Drinkwater}}{{Bekki}
  et~al.}{2001}]{bekki01}
{Bekki} K.,  {Couch} W.~J.,    {Drinkwater} M.~J.,  2001, ApJL, 552, L105

\bibitem[\protect\citeauthoryear{{Bertin} \& {Arnouts}}{{Bertin} \&
  {Arnouts}}{1996}]{bertin}
{Bertin} E.,  {Arnouts} S.,  1996, A\&AS, 117, 393

\bibitem[\protect\citeauthoryear{{Brodie}, {Romanowsky}, {Strader} \&
  {Forbes}}{{Brodie} et~al.}{2011}]{brodie11}
{Brodie} J.~P.,  {Romanowsky} A.~J.,  {Strader} J.,    {Forbes} D.~A.,  2011,
  ArXiv e-prints

\bibitem[\protect\citeauthoryear{{Brodie}, {Schroder}, {Huchra}, {Phillips},
  {Kissler-Patig} \& {Forbes}}{{Brodie} et~al.}{1998}]{brodie98}
{Brodie} J.~P.,  {Schroder} L.~L.,  {Huchra} J.~P.,  {Phillips} A.~C.,
  {Kissler-Patig} M.,    {Forbes} D.~A.,  1998, AJ, 116, 691

\bibitem[\protect\citeauthoryear{{Brough}, {Tran}, {Sharp}, {von der Linden} \&
  {Couch}}{{Brough} et~al.}{2011}]{brough11}
{Brough} S.,  {Tran} K.-V.,  {Sharp} R.~G.,  {von der Linden} A.,    {Couch}
  W.~J.,  2011, MNRAS, 414, L80

\bibitem[\protect\citeauthoryear{{Br{\"u}ns}, {Kroupa}, {Fellhauer}, {Metz} \&
  {Assmann}}{{Br{\"u}ns} et~al.}{2011}]{bruns11}
{Br{\"u}ns} R.~C.,  {Kroupa} P.,  {Fellhauer} M.,  {Metz} M.,    {Assmann} P.,
  2011, A\&A, 529, A138

\bibitem[\protect\citeauthoryear{{Bruzual} \& {Charlot}}{{Bruzual} \&
  {Charlot}}{2003}]{bruzualcharlot03}
{Bruzual} G.,  {Charlot} S.,  2003, MNRAS, 344, 1000

\bibitem[\protect\citeauthoryear{{Burstein} \& {Heiles}}{{Burstein} \&
  {Heiles}}{1984}]{burstein84}
{Burstein} D.,  {Heiles} C.,  1984, ApJS, 54, 33

\bibitem[\protect\citeauthoryear{{Canning}, {Fabian}, {Johnstone}, {Sanders},
  {Conselice}, {Crawford}, {Gallagher} \& {Zweibel}}{{Canning}
  et~al.}{2010}]{canning10}
{Canning} R.~E.~A.,  {Fabian} A.~C.,  {Johnstone} R.~M.,  {Sanders} J.~S.,
  {Conselice} C.~J.,  {Crawford} C.~S.,  {Gallagher} J.~S.,    {Zweibel} E.,
  2010, MNRAS, 405, 115

\bibitem[\protect\citeauthoryear{{Chiboucas}, {Tully}, {Marzke}, {Phillipps},
  {Price}, {Peng}, {Trentham}, {Carter} \& {Hammer}}{{Chiboucas}
  et~al.}{2011}]{chiboucas11}
{Chiboucas} K.,  {Tully} R.~B.,  {Marzke} R.~O.,  {Phillipps} S.,  {Price} J.,
  {Peng} E.~W.,  {Trentham} N.,  {Carter} D.,    {Hammer} D.,  2011, ApJ, 737,
  86

\bibitem[\protect\citeauthoryear{{Conselice}, {Gallagher} III \&
  {Wyse}}{{Conselice} et~al.}{2001}]{conselice1275}
{Conselice} C.~J.,  {Gallagher} III J.~S.,    {Wyse} R.~F.~G.,  2001, AJ, 122,
  2281

\bibitem[\protect\citeauthoryear{{Conselice}, {Gallagher} III \&
  {Wyse}}{{Conselice} et~al.}{2003}]{cgw03}
{Conselice} C.~J.,  {Gallagher} III J.~S.,    {Wyse} R.~F.~G.,  2003, AJ, 125,
  66

\bibitem[\protect\citeauthoryear{{de Rijcke}, {Penny}, {Conselice}, {Valcke} \&
  {Held}}{{de Rijcke} et~al.}{2009}]{derijcke09}
{de Rijcke} S.,  {Penny} S.~J.,  {Conselice} C.~J.,  {Valcke} S.,    {Held}
  E.~V.,  2009, MNRAS, 393, 798

\bibitem[\protect\citeauthoryear{{de Vaucouleurs}, {de Vaucouleurs}, {Corwin}
  Jr., {Buta}, {Paturel} \& {Fouque}}{{de Vaucouleurs} et~al.}{1991}]{refcat}
{de Vaucouleurs} G.,  {de Vaucouleurs} A.,  {Corwin} Jr. H.~G.,  {Buta} R.~J.,
  {Paturel} G.,    {Fouque} P.,  1991, {Third Reference Catalogue of Bright
  Galaxies}

\bibitem[\protect\citeauthoryear{{Drinkwater}, {Gregg}, {Hilker}, {Bekki},
  {Couch}, {Ferguson}, {Jones} \& {Phillipps}}{{Drinkwater}
  et~al.}{2003}]{drinkwater03}
{Drinkwater} M.~J.,  {Gregg} M.~D.,  {Hilker} M.,  {Bekki} K.,  {Couch} W.~J.,
  {Ferguson} H.~C.,  {Jones} J.~B.,    {Phillipps} S.,  2003, Nat, 423, 519

\bibitem[\protect\citeauthoryear{{Drinkwater}, {Jones}, {Gregg} \&
  {Phillipps}}{{Drinkwater} et~al.}{2000}]{drinkwater00}
{Drinkwater} M.~J.,  {Jones} J.~B.,  {Gregg} M.~D.,    {Phillipps} S.,  2000,
  PASA, 17, 227

\bibitem[\protect\citeauthoryear{{Evstigneeva}, {Drinkwater}, {Peng}, {Hilker},
  {De Propris}, {Jones}, {Phillipps}, {Gregg} \& {Karick}}{{Evstigneeva}
  et~al.}{2008}]{evstigneeva08}
{Evstigneeva} E.~A.,  {Drinkwater} M.~J.,  {Peng} C.~Y.,  {Hilker} M.,  {De
  Propris} R.,  {Jones} J.~B.,  {Phillipps} S.,  {Gregg} M.~D.,    {Karick}
  A.~M.,  2008, AJ, 136, 461

\bibitem[\protect\citeauthoryear{{Evstigneeva}, {Gregg}, {Drinkwater} \&
  {Hilker}}{{Evstigneeva} et~al.}{2007}]{evstigneeva07}
{Evstigneeva} E.~A.,  {Gregg} M.~D.,  {Drinkwater} M.~J.,    {Hilker} M.,
  2007, AJ, 133, 1722

\bibitem[\protect\citeauthoryear{{Fellhauer} \& {Kroupa}}{{Fellhauer} \&
  {Kroupa}}{2002}]{fellhauer02}
{Fellhauer} M.,  {Kroupa} P.,  2002, MNRAS, 330, 642

\bibitem[\protect\citeauthoryear{{Forbes} \& {Kroupa}}{{Forbes} \&
  {Kroupa}}{2011}]{forbes11b}
{Forbes} D.~A.,  {Kroupa} P.,  2011, PASA, 28, 77

\bibitem[\protect\citeauthoryear{{Forbes}, {Spitler}, {Strader}, {Romanowsky},
  {Brodie} \& {Foster}}{{Forbes} et~al.}{2011}]{forbes11}
{Forbes} D.~A.,  {Spitler} L.~R.,  {Strader} J.,  {Romanowsky} A.~J.,  {Brodie}
  J.~P.,    {Foster} C.,  2011, MNRAS, 413, 2943

\bibitem[\protect\citeauthoryear{{Fukugita}, {Shimasaku} \&
  {Ichikawa}}{{Fukugita} et~al.}{1995}]{fukugita95}
{Fukugita} M.,  {Shimasaku} K.,    {Ichikawa} T.,  1995, PASP, 107, 945

\bibitem[\protect\citeauthoryear{{Gilmore}, {Wilkinson}, {Wyse}, {Kleyna},
  {Koch}, {Evans} \& {Grebel}}{{Gilmore} et~al.}{2007}]{gilmore07}
{Gilmore} G.,  {Wilkinson} M.~I.,  {Wyse} R.~F.~G.,  {Kleyna} J.~T.,  {Koch}
  A.,  {Evans} N.~W.,    {Grebel} E.~K.,  2007, ApJ, 663, 948

\bibitem[\protect\citeauthoryear{{Graham} \& {Guzm{\'a}n}}{{Graham} \&
  {Guzm{\'a}n}}{2003}]{graham03}
{Graham} A.~W.,  {Guzm{\'a}n} R.,  2003, AJ, 125, 2936

\bibitem[\protect\citeauthoryear{{Grebel}, {Gallagher} III \&
  {Harbeck}}{{Grebel} et~al.}{2003}]{grebel03}
{Grebel} E.~K.,  {Gallagher} III J.~S.,    {Harbeck} D.,  2003, AJ, 125, 1926

\bibitem[\protect\citeauthoryear{{Gregg}, {Drinkwater}, {Evstigneeva}, {Jurek},
  {Karick}, {Phillipps}, {Bridges}, {Jones}, {Bekki} \& {Couch}}{{Gregg}
  et~al.}{2009}]{gregg09}
{Gregg} M.~D.,  {Drinkwater} M.~J.,  {Evstigneeva} E.,  {Jurek} R.,  {Karick}
  A.~M.,  {Phillipps} S.,  {Bridges} T.,  {Jones} J.~B.,  {Bekki} K.,
  {Couch} W.~J.,  2009, AJ, 137, 498

\bibitem[\protect\citeauthoryear{{Ha{\c s}egan}, {Jord{\'a}n}, {C{\^o}t{\'e}},
  {Djorgovski}, {McLaughlin}, {Blakeslee}, {Mei}, {West}, {Peng}, {Ferrarese},
  {Milosavljevi{\'c}}, {Tonry} \& {Merritt}}{{Ha{\c s}egan}
  et~al.}{2005}]{hasegan05}
{Ha{\c s}egan} M.,  {Jord{\'a}n} A.,  {C{\^o}t{\'e}} P.,  {Djorgovski} S.~G.,
  {McLaughlin} D.~E.,  {Blakeslee} J.~P.,  {Mei} S.,  {West} M.~J.,  {Peng}
  E.~W.,  {Ferrarese} L.,  {Milosavljevi{\'c}} M.,  {Tonry} J.~L.,    {Merritt}
  D.,  2005, ApJ, 627, 203

\bibitem[\protect\citeauthoryear{{Harris}}{{Harris}}{2009a}]{harris09}
{Harris} W.~E.,  2009a, ApJ, 699, 254

\bibitem[\protect\citeauthoryear{{Harris}}{{Harris}}{2009b}]{harris09b}
{Harris} W.~E.,  2009b, ApJ, 699, 254

\bibitem[\protect\citeauthoryear{{Hilker}, {Infante}, {Vieira}, {Kissler-Patig}
  \& {Richtler}}{{Hilker} et~al.}{1999}]{hilker99}
{Hilker} M.,  {Infante} L.,  {Vieira} G.,  {Kissler-Patig} M.,    {Richtler}
  T.,  1999, A\&AS, 134, 75

\bibitem[\protect\citeauthoryear{{Holtzman}, {Faber}, {Shaya}, {Lauer},
  {Groth}, {Hunter}, {Baum}, {Ewald}, {Hester}, {Light}, {Lynds}, {O'Neil} Jr.
  \& {Westphal}}{{Holtzman} et~al.}{1992}]{holtzman92}
{Holtzman} J.~A.,  {Faber} S.~M.,  {Shaya} E.~J.,  {Lauer} T.~R.,  {Groth} J.,
  {Hunter} D.~A.,  {Baum} W.~A.,  {Ewald} S.~P.,  {Hester} J.~J.,  {Light}
  R.~M.,  {Lynds} C.~R.,  {O'Neil} Jr. E.~J.,    {Westphal} J.~A.,  1992, AJ,
  103, 691

\bibitem[\protect\citeauthoryear{{Huxor}, {Phillipps}, {Price} \&
  {Harniman}}{{Huxor} et~al.}{2011}]{huxor11}
{Huxor} A.~P.,  {Phillipps} S.,  {Price} J.,    {Harniman} R.,  2011, MNRAS,
  414, 3557

\bibitem[\protect\citeauthoryear{{Kent}}{{Kent}}{1987}]{kent87}
{Kent} S.~M.,  1987, AJ, 94, 306

\bibitem[\protect\citeauthoryear{{Kissler-Patig}, {Jord{\'a}n} \&
  {Bastian}}{{Kissler-Patig} et~al.}{2006}]{kisslerpatig06}
{Kissler-Patig} M.,  {Jord{\'a}n} A.,    {Bastian} N.,  2006, A\&A, 448, 1031

\bibitem[\protect\citeauthoryear{{Larsen}}{{Larsen}}{1999}]{larsen09}
{Larsen} S.~S.,  1999, A\&AS, 139, 393

\bibitem[\protect\citeauthoryear{{Madrid}, {Graham}, {Harris}, {Goudfrooij},
  {Forbes}, {Carter}, {Blakeslee}, {Spitler} \& {Ferguson}}{{Madrid}
  et~al.}{2010}]{madrid10}
{Madrid} J.~P.,  {Graham} A.~W.,  {Harris} W.~E.,  {Goudfrooij} P.,  {Forbes}
  D.~A.,  {Carter} D.,  {Blakeslee} J.~P.,  {Spitler} L.~R.,    {Ferguson}
  H.~C.,  2010, ApJ, 722, 1707

\bibitem[\protect\citeauthoryear{{Martin}, {de Jong} \& {Rix}}{{Martin}
  et~al.}{2008}]{martin08}
{Martin} N.~F.,  {de Jong} J.~T.~A.,    {Rix} H.-W.,  2008, ApJ, 684, 1075

\bibitem[\protect\citeauthoryear{{Mieske}, {Hilker}, {Jord{\'a}n}, {Infante},
  {Kissler-Patig}, {Rejkuba}, {Richtler}, {C{\^o}t{\'e}}, {Baumgardt}, {West},
  {Ferrarese} \& {Peng}}{{Mieske} et~al.}{2008}]{mieske08}
{Mieske} S.,  {Hilker} M.,  {Jord{\'a}n} A.,  {Infante} L.,  {Kissler-Patig}
  M.,  {Rejkuba} M.,  {Richtler} T.,  {C{\^o}t{\'e}} P.,  {Baumgardt} H.,
  {West} M.~J.,  {Ferrarese} L.,    {Peng} E.~W.,  2008, A\&A, 487, 921

\bibitem[\protect\citeauthoryear{{Mieske}, {Hilker}, {Misgeld}, {Jord{\'a}n},
  {Infante} \& {Kissler-Patig}}{{Mieske} et~al.}{2009}]{mieske09}
{Mieske} S.,  {Hilker} M.,  {Misgeld} I.,  {Jord{\'a}n} A.,  {Infante} L.,
  {Kissler-Patig} M.,  2009, A\&A, 498, 705

\bibitem[\protect\citeauthoryear{{Minkowski}}{{Minkowski}}{1957}]{minkowski57}
{Minkowski} R.,  1957, in {H.~C.~van de Hulst} ed., Radio astronomy Vol.~4 of
  IAU Symposium, {Optical investigations of radio sources (Introductory
  Lecture)}.
pp 107--+

\bibitem[\protect\citeauthoryear{{Misgeld} \& {Hilker}}{{Misgeld} \&
  {Hilker}}{2011}]{misgeld11a}
{Misgeld} I.,  {Hilker} M.,  2011, MNRAS, 414, 3699

\bibitem[\protect\citeauthoryear{{Misgeld}, {Hilker} \& {Mieske}}{{Misgeld}
  et~al.}{2009}]{misgeld09}
{Misgeld} I.,  {Hilker} M.,    {Mieske} S.,  2009, VizieR Online Data Catalog,
  349, 60683

\bibitem[\protect\citeauthoryear{{Misgeld}, {Mieske} \& {Hilker}}{{Misgeld}
  et~al.}{2008}]{misgeld08}
{Misgeld} I.,  {Mieske} S.,    {Hilker} M.,  2008, astro-ph, 486, 697

\bibitem[\protect\citeauthoryear{{Misgeld}, {Mieske}, {Hilker}, {Richtler},
  {Georgiev} \& {Schuberth}}{{Misgeld} et~al.}{2011}]{misgeld11b}
{Misgeld} I.,  {Mieske} S.,  {Hilker} M.,  {Richtler} T.,  {Georgiev} I.~Y.,
  {Schuberth} Y.,  2011, A\&A, 531, A4+

\bibitem[\protect\citeauthoryear{{Norris} \& {Kannappan}}{{Norris} \&
  {Kannappan}}{2011}]{norris11}
{Norris} M.~A.,  {Kannappan} S.~J.,  2011, MNRAS, 414, 739

\bibitem[\protect\citeauthoryear{{Penny} \& {Conselice}}{{Penny} \&
  {Conselice}}{2008}]{penny08}
{Penny} S.~J.,  {Conselice} C.~J.,  2008, MNRAS, 383, 247

\bibitem[\protect\citeauthoryear{{Penny}, {Conselice}, {de Rijcke} \&
  {Held}}{{Penny} et~al.}{2009}]{penny09}
{Penny} S.~J.,  {Conselice} C.~J.,  {de Rijcke} S.,    {Held} E.~V.,  2009,
  MNRAS, 393, 1054

\bibitem[\protect\citeauthoryear{{Price}, {Phillipps}, {Huxor}, {Trentham},
  {Ferguson}, {Marzke}, {Hornschemeier}, {Goudfrooij}, {Hammer}, {Tully},
  {Chiboucas}, {Smith}, {Carter}, {Merritt}, {Balcells}, {Erwin} \&
  {Puzia}}{{Price} et~al.}{2009}]{price09}
{Price} J.,  {Phillipps} S.,  {Huxor} A.,  {Trentham} N.,  {Ferguson} H.~C.,
  {Marzke} R.~O.,  {Hornschemeier} A.,  {Goudfrooij} P.,  {Hammer} D.,  {Tully}
  R.~B.,  {Chiboucas} K.,  {Smith} R.~J.,  {Carter} D.,  {Merritt} D.,
  {Balcells} M.,  {Erwin} P.,    {Puzia} T.~H.,  2009, MNRAS, 397, 1816

\bibitem[\protect\citeauthoryear{{Schlegel}, {Finkbeiner} \&
  {Davis}}{{Schlegel} et~al.}{1998}]{schlegel}
{Schlegel} D.~J.,  {Finkbeiner} D.~P.,    {Davis} M.,  1998, ApJ, 500, 525

\bibitem[\protect\citeauthoryear{{Sirianni}, {Jee}, {Ben{\'{\i}}tez},
  {Blakeslee}, {Martel}, {Meurer}, {Clampin}, {De Marchi}, {Ford}, {Gilliland},
  {Hartig}, {Illingworth}, {Mack} \& {McCann}}{{Sirianni}
  et~al.}{2005}]{sirianni}
{Sirianni} M.,  {Jee} M.~J.,  {Ben{\'{\i}}tez} N.,  {Blakeslee} J.~P.,
  {Martel} A.~R.,  {Meurer} G.,  {Clampin} M.,  {De Marchi} G.,  {Ford} H.~C.,
  {Gilliland} R.,  {Hartig} G.~F.,  {Illingworth} G.~D.,  {Mack} J.,
  {McCann} W.~J.,  2005, PASP, 117, 1049

\bibitem[\protect\citeauthoryear{{Smith Castelli}, {Faifer}, {Richtler} \&
  {Bassino}}{{Smith Castelli} et~al.}{2008}]{smithcastelli08}
{Smith Castelli} A.~V.,  {Faifer} F.~R.,  {Richtler} T.,    {Bassino} L.~P.,
  2008, MNRAS, 391, 685

\bibitem[\protect\citeauthoryear{{Strader} \& {Smith}}{{Strader} \&
  {Smith}}{2008}]{strader08}
{Strader} J.,  {Smith} G.~H.,  2008, AJ, 136, 1828

\bibitem[\protect\citeauthoryear{{Struble} \& {Rood}}{{Struble} \&
  {Rood}}{1999}]{Stublerood99}
{Struble} M.~F.,  {Rood} H.~J.,  1999, ApJS, 125, 35

\bibitem[\protect\citeauthoryear{{Wehner}, {Harris}, {Whitmore}, {Rothberg} \&
  {Woodley}}{{Wehner} et~al.}{2008}]{wehner08}
{Wehner} E.~M.~H.,  {Harris} W.~E.,  {Whitmore} B.~C.,  {Rothberg} B.,
  {Woodley} K.~A.,  2008, ApJ, 681, 1233

\end{thebibliography}
\end{document}